\documentclass{sig-alternate}
\usepackage{mathptmx}

\pdfoutput=1

\usepackage{cite}
\usepackage{amsmath,amssymb,amsfonts}
\usepackage{graphicx}
\usepackage{textcomp}
\usepackage{color,xcolor}
\usepackage{fancyhdr}
\usepackage[hyphens]{url}
\usepackage[normalem]{ulem}
\usepackage[final]{microtype}
\usepackage[keeplastbox]{flushend}
\usepackage[final]{microtype}
\usepackage{subcaption}
\usepackage{balance}
\usepackage[inline]{enumitem}
\usepackage{tikz}
\newcommand*\circled[1]{\tikz[baseline=(char.base)]{
            \node[shape=circle,draw,inner sep=1pt] (char) {\footnotesize #1};}}
\usepackage[bookmarks=true,breaklinks=true,letterpaper=true,colorlinks,linkcolor=black,citecolor=blue,urlcolor=black]{hyperref}

\newcommand{\ignore}[1]{}

\makeatletter
\newcommand\fakeheader[1]{\noindent\textbf{#1: }%
\@ifnextchar\par{\@gobble}{}}
\makeatother

\newcommand\cs[1]{}
\newcommand\hms[1]{}
\newcommand\sk[1]{}

\def\BibTeX{{\rm B\kern-.05em{\sc i\kern-.025em b}\kern-.08em
    T\kern-.1667em\lower.7ex\hbox{E}\kern-.125emX}}

\title{Selectively Delaying Instructions to Prevent Microarchitectural Replay Attacks}
\numberofauthors{3} 
\author{
  \alignauthor
  Christos Sakalis\\
    \affaddr{Uppsala University, Sweden}\\
    \email{\normalsize{christos.sakalis@it.uu.se}}
  \alignauthor
  Stefanos Kaxiras\\
    \affaddr{Uppsala University, Sweden}\\
    \email{\normalsize{stefanos.kaxiras@it.uu.se}}
  \alignauthor
  Magnus Sj\"alander\\
    \affaddr{Norwegian University of Science and Technology}\\
    \email{\normalsize{magnus.sjalander@ntnu.no}}
  }
\begin{document}
\maketitle

% \author{Christos Sakalis}
% \orcid{0000-0003-4172-8607}
% \affiliation{%
  % \institution{Uppsala University}
  % \streetaddress{Box 337}
  % \city{Uppsala}
  % \postcode{75105}
  % \country{Sweden}}
% \email{christos.sakalis@it.uu.se}
% \author{Stefanos Kaxiras}
% \orcid{0000-0001-8267-0232}
% \affiliation{%
  % \institution{Uppsala University}
  % \streetaddress{Box 337}
  % \city{Uppsala}
  % \postcode{75105}
  % \country{Sweden}}
% \email{stefanos.kaxiras@it.uu.se}
% \author{Magnus Sj\"alander}
% \orcid{0000-0003-4232-6976}
% \affiliation{%
  % \institution{Norwegian University of Science and Technology}
  % \streetaddress{Sem S{\ae}landsvei 9}
  % \postcode{7491}
  % \city{Trondheim}
  % \country{Norway}
% }
% \email{magnus.sjalander@ntnu.no}
% \author{C. Sakalis, S. Kaxiras, M. Sj\"alander}

% \renewcommand{\shortauthors}{C. Sakalis, S. Kaxiras, and M. Sj\"alander}

\begin{abstract}

  MicroScope, and microarchitectural replay attacks in general, take advantage of
  the characteristics of speculative execution to trap the execution of the
  victim application in an infinite loop, enabling the attacker to amplify a
  side-channel attack by executing it indefinitely.  %
  Due to the nature of the replay, it can be used to effectively attack security
  critical trusted execution environments (secure enclaves), even under
  conditions where a side-channel attack would not be possible. %
  At the same time, unlike speculative side-channel attacks, MicroScope can be
  used to amplify the correct path of execution, rendering many existing
  speculative side-channel defences ineffective. %

  In this work, we generalize microarchitectural replay attacks beyond
  MicroScope and present an efficient defence against them. %
  We make the observation that such attacks rely on repeated squashes of
  so-called ``replay handles'' and that the instructions causing the
  side-channel must reside in the same reorder buffer window as the handles. %
  We propose Delay-on-Squash, a technique for tracking squashed instructions and
  preventing them from being replayed by speculative replay handles. %
  Our evaluation shows that it is possible to achieve full security against
  microarchitectural replay attacks with very modest hardware requirements,
  while still maintaining 97\% of the insecure baseline performance. %

  \cs{TODO: Try to add a potential attack using the new handles.}

\end{abstract}

  % <concept_id>10002978.10003001.10010777.10011702</concept_id>
  % <concept_desc>Security and privacy~Side-channel analysis and countermeasures</concept_desc>

% \ccsdesc[500]{Security and privacy~Side-channel analysis and countermeasures}

% \keywords{speculative execution, side-channel attacks, caches}

\maketitle

\section{Introduction}
\label{sec:introduction}

% \cs{1. No computer system is absolutely secure.}
% \cs{2. Security comes at a cost.}
% \cs{filter width vs number of filters}

% \sk{Make sure that you explain how we loose performance.}

With the ever rising concern regarding privacy and security in the modern digital
world, we have seen an increased interest in hardware trusted execution
environments (also known as \textbf{secure enclaves}) from
computer architects, software developers, and security researchers. %
Such enclaves provide hardware enforced security guarantees that protect the
enclaved code from outside interference, including interference from the
operating system (OS) or untrusted hardware found in the system. %
This is achieved through a variety of security measures such as memory
encryption and attestation. %
These measures come at a performance (and energy) cost, so enclaves are only
typically used for applications where security, instead of efficiency, is the
paramount concern. %

Unfortunately, even though enclaves are able to prevent malicious system
components from directly interfering with the enclaved code, there are still
numerous attacks that are made possible through side-channels. %
By exploiting architectural and microarchitectural behavior of the system in
unintended ways, these attacks (e.g., Zombieload~\cite{schwarz_zombieload_2019} or
TLBleed~\cite{gras_tlbleak_2018}) create covert communication channels and leak
secret information. %
Thankfully, enclaves have remained effective, as side-channels are typically
very noisy communication channels and often require several iterations of the
same attack before being able to reliably leak any information. %
While it is not hard to imagine cases where the attacker is targeting specific
immutable data and the enclaved code can be arbitrarily triggered (e.g., to
encrypt some data), in a lot of cases (e.g., SGX implementations of
Tor~\cite{WWWtor,kim_sgx-tor_2018}, secure database implementations, or systems
secured against rollbacks~\cite{matetic+:SECURITY2017rote}) the attacker targets
transient execution data (e.g., Tor traffic) and only has one opportunity to
perform the attack and leak information. %
In these cases, the majority of the available side-channels are not effective,
as it is not possible to distinguish a single iteration of the attack from
system noise~\cite{lyu_survey_2018, aldaya_port_2019, maurice_hello_2017}. %

MicroScope, and \emph{microarchitectural replay attacks} in general, as
introduced by Skarlatos et al.~\cite{skarlatos_microscope:_2019},
enable an attacker to ``trap'' the execution of an enclaved application and
force it to re-execute specific regions of code ad infinitum. %
With MicroScope, this is achieved by abusing a combination of speculative
execution and page fault handling, in cases where the latter is still delegated
to the (malicious/compromised) OS. %
Under typical execution, regardless if the code is executed in an enclave or
not, if an address translation misses in the translation lookaside buffer (TLB)
a page table walk is triggered. %
While the page walk is happening, the application is able to continue executing
speculatively, as long as the instructions do not depend on the faulting memory
instruction. %
If during the page walk it is determined that the page is not available and that
the OS needs to be invoked, then the speculatively executed instructions are
squashed and execution restarts from the faulting instruction. %
This time another page walk might be needed, as the translation still does not
exist in the TLB, but since the operating system has now mapped the page, the
page walk typically succeeds. %
MicroScope takes advantage of this behavior by having the OS signal that it has
mapped the page without actually doing so. %
This traps the victim application in a loop where a memory instruction (referred
to as \textbf{``the handle''}) misses in the TLB, triggers a page walk and
continues executing speculatively, triggers a page fault, squashes the
speculatively executed instructions, re-executes the faulting instruction and
misses in the TLB again. %
By triggering these loops at specific parts of the code, just before the
instructions that cause the side-channel information leakage, the attacker can
repeat the side-channel until all the underlying noise is filtered out, making
even the least reliable side-channels easy to exploit. %

While MicroScope focuses specifically on abusing the page handling mechanism
found on some enclaves, other sources of speculation and re-execution can be
similarly abused. %
For this reason, in this work we aim to solve the problem of microarchitectural
replay attacks in general, and not just MicroScope. %
Specifically:

\begin{itemize}

    \item We expand from re-execution brought on by page faults to re-execution
      brought on by any form of speculation in modern processors, some of which
      do not require a malicious OS. %

    \item We also extend the method developed by Skarlatos et al. to be
      applicable to cases where a single handle cannot by itself trigger a large
      number of re-executions, by introducing a method that utilizes multiple
      handles. %

    \item Finally, we introduce \textbf{Delay-on-Squash}, a solution to this
      critical issue of microarchitectural replay attacks, which can
      transparently provide protection for MicroScope and future attacks, while
      avoiding significant impact on performance, energy, area, and
      implementation complexity. %

\end{itemize}

% {\color{red} STEFANOS: THIS IS CRITICAL: You must argue, here, either at the
% beginning or the end of the following paragraph, why a naive solution is not
% enough.  LATER ON YOU SAY: ``In addition, if the system restricted the number of
% times each instruction is allowed to misspeculate and then be re-executed
% speculatively, e.g., as a simple but naive solution to MicroScope, an attacker
% would still be able to use multiple handles to force a finite but tangible
% number of replays''. YOU MUST ALSO SAY SOMETHING ABOUT PERFORMANCE OF THE NAIVE
% SOLUTION }
% Chris: The performance of the naive solution would be awesome
% \cs{only being run in enclave, don't run hpc code in enclaves}

% \fakeheader{The need for a new solution}
A very simple but na\"ive solution would be to simply disallow speculative
execution after a page walk miss, but this would only protect against MicroScope
itself and not against microarchitectural replay attacks that utilize other
handle types or multiple handles (\autoref{sec:beyond-microscope}). %
At the same time, we cannot just disable speculation every time a squash
happens, as that would be detrimental to the performance of the system
(\autoref{sec:evaluation}). %
Furthermore, the purpose of a replay attack (primarily in secure enclaves) is to
amplify side-channel instructions. %
Speculative side-channel defences that track the \emph{data dependencies} of
side-channel instructions, such as NDA~\cite{weisse_nda:_2019},
STT~\cite{yu_speculative_2019, sdo20}, and others~\cite{barber_specshield_2019,
conditional19, fustos_spectreguard:_2019, barber20}, do not work in
this context, because the side-channel instructions may actually be in the
\emph{correct path of execution} and \emph{can also be fed with non-speculative
data coming from before the point of misspeculation} that is used for replay. %
Broader defences (not restricted to data dependencies) that could work in such a
case, e.g., InvisiSpec~\cite{yan_invisispec:MICRO2018},
Delay-on-Miss~\cite{dom19, dom20}, and many others~\cite{ghosts, safespec_2019,
Saileshwar19, revice20, muontrap20, kiriansky_dawg:MICRO2018,
taram_context-sensitive_2019}, not only focus on a small subset of side-channels
but also incur much heavier penalties. %
Our goal is to propose a highly-efficient new defence to effectively prevent
replay attacks with minimum performance cost and hardware overhead. %

Delay-on-Squash selectively delays the speculative execution of instructions
when it detects that they might be used as part of a microarchitectural replay
attack. %
We observe that if an attack requires microarchitectural replay to be
successful, then we do not need to restrict all speculation but rather only
repeating speculation interleaved with misspeculation. %
To achieve this, we use a lightweight mechanism  that tracks (using Bloom
filters) which instructions have been squashed and later re-issued under unsafe
conditions, and prevents them from executing speculatively. %
Our evaluation shows that a fully secure configuration of Delay-on-Squash,
requiring little storage and overall overhead, can achieve $97\%$ of the
performance of a baseline insecure out-of-order CPU. %

\section{Side-channels}
\label{sec:side-channels}

In order to provide compatibility across different hardware implementations,
modern CPU architectures separate the visible \emph{architectural} behavior and
state of the system from the underlying \emph{microarchitectural}
implementation. %
For each visible architectural state, there exist one or more corresponding
hidden microarchitectural states ($\mu$-states)~\cite{mcilroy_spectre_2019}. %
The users/applications interact with the architectural state but have limited or
no access to the underlying $\mu$-state. %
However, while it is usually not possible to observe the $\mu$-state directly,
it is possible to infer it based on observable side-effects. %

% \cs{
% For example, while it is generally not possible for a program to query the data
% cache directly, it is possible to determine if a cache line exists in the cache
% by simply measuring the amount of time required for loading the data. %
% Similarly, a program is not able to query the availability of the functional
% units (FUs), but it is possible to infer FU contention based on the execution
% time of instructions. %
% Other than timing, other side-effects are also observable, such as energy
% consumption or EF radiation. %
% }

Microarchitectural side-channel attacks take advantage of the $\mu$-state of
modern CPUs to %transfer and
leak information under conditions where it is not
possible to do so on the architectural level. %
For example, cache side-channels take advantage of the difference in timing
between a hit or a miss in the cache to encode
information~\cite{osvik_cache_2006, yarom_flush+_2014, lyu_survey_2018}, by
indirectly manipulating and probing the state of the cache through normal memory
operations. %
Similar timing side-channels can also be constructed in other parts of the
system~\cite{ge_survey_2018}, such as by utilizing functional unit (FU)
contention. %
Finally, non-timing side-channels are also possible, exploiting side-effects of
the execution such as power consumption~\cite{kocher+:CRYPTO1999dpa} or EMF
radiation~\cite{gandolfi+2001emf}. %

As these side-channels are not purposely designed communication channels but
rather side-effects of the normal architectural and microarchitectural behavior
of the system, they are inherently noisy and unreliable. %
For example, when using a cache-based side-channel, there is nothing to prevent
the cacheline(s) being used for the side-channel from being evicted by a third
process in the system. %
Similarly, interrupts, context switches, and other interruptions in the
application execution can also disrupt the side-channel. %
Since the system does not provide any architectural mechanisms for synchronizing
the transmitter and the receiver during side-channel operations, these also have
to be constructed using the side-channel itself or other mechanisms. %

All these issues make exploiting side-channels harder, but not impossible. %
After all, these issues can be found in conventional communication
channels as well, especially in the layers closest to the actual hardware and
transmission mediums. %
Similarly to how modern communication protocols are designed with the
underlying channel characteristics in mind, so can protocols for side-channels
be designed. %
For example, noise on a side-channel can be filtered out using error detection
and correction codes, combined with statistical methods. %
Maurice et al.~\cite{maurice_hello_2017} have presented one such protocol where
they were able to implement \texttt{ssh} communications over a cache-based
side-channel. %
The question then becomes \emph{not if a side-channel can be exploited but
rather how easy, reliable, and fast the channel is}. %
As many of the underlying issues can be resolved by simply repeating the
transmission of information through the side-channel until successful, whether a
side-channel can be \emph{practically} exploited becomes a function of the delay
between each retransmission (i.e., how fast can the side-channel be repeated)
and the average number of retransmissions necessary (i.e., how many times does
the side-channel have to be repeated). %

Under the conditions described by Maurice et al., both the transmitter and the
receiver are under the full control of the attacker. %
This enables the attacker to not only control when the side-channel transmission
takes place, to better synchronize the transmitter and the receiver, but to also
repeat the transmission as many times as necessary. %
However, this is not always the case. %
For example, sometimes the transmitter is not a purposely designed application
but a targeted victim, such as a cryptographic application running in a secure
enclave. %
The attacker then uses side-channels to monitor the behavior of the application
under normal execution and infer sensitive information, such as cryptographic
keys. %
In some cases, the attacker is able to either directly execute or trigger the
execution of the victim application at will, repeating the execution as many
times as necessary to extract the keys. %
This ability to \emph{replay} the victim code multiple times is crucial in being
able to reliably exploit the utilized side-channel, due to the issues we have
already discussed. %
This is where MicroScope, a groundbreaking \emph{microarchitectural replay
attack} developed by Skarlatos et al.~\cite{skarlatos_microscope:_2019} comes
into play. %

\section{Microarchitectural Replay Attacks}
\label{sec:mra}

% \cs{Programs are expected to handle side-channels by themselves in SGX, but
% MicroScope takes away the control from the program. We can restore back that
% control with a tiny performance penalty.}

Microarchitectural replay attacks, as introduced by Skarlatos et
al.~\cite{skarlatos_microscope:_2019}, are not by themselves a side-channel
attack, speculative or otherwise. %
Instead, they can be seen as a tool to \emph{amplify the effects of
side-channel attacks}, enabling the attacker to mount a successful attack under
conditions where it would not be possible otherwise. %
This is why microarchitectural replay can be so dangerous: even the smallest,
most innocuous amount of information leakage can be amplified and
abused. %
This is particularly dangerous when applied to secure enclaves, where the
applications are security sensitive and the programmers themselves are typically expected
to manage the risk of side-channels. %
In addition, even though they exploit speculative execution, microarchitectural
replay attacks \emph{are not the same} as speculative side-channel attacks. %
Whereas the latter target the wrong execution path, effectively bypassing
software and hardware barriers to access information illegally,
microarchitectural replay attacks can amplify even the \emph{correct} path of
execution. %
This means that defences that stop speculative data
transmission~\cite{weisse_nda:_2019, yu_speculative_2019, sdo20} are irrelevant
since a replay attack can also amplify side-channel instructions that are on the
correct path. %

\subsection{MicroScope}
\label{sec:microscope}

\begin{figure*}[t]
  \centering
  \begin{subfigure}[t]{0.2960\textwidth}
    \includegraphics[width=\textwidth]{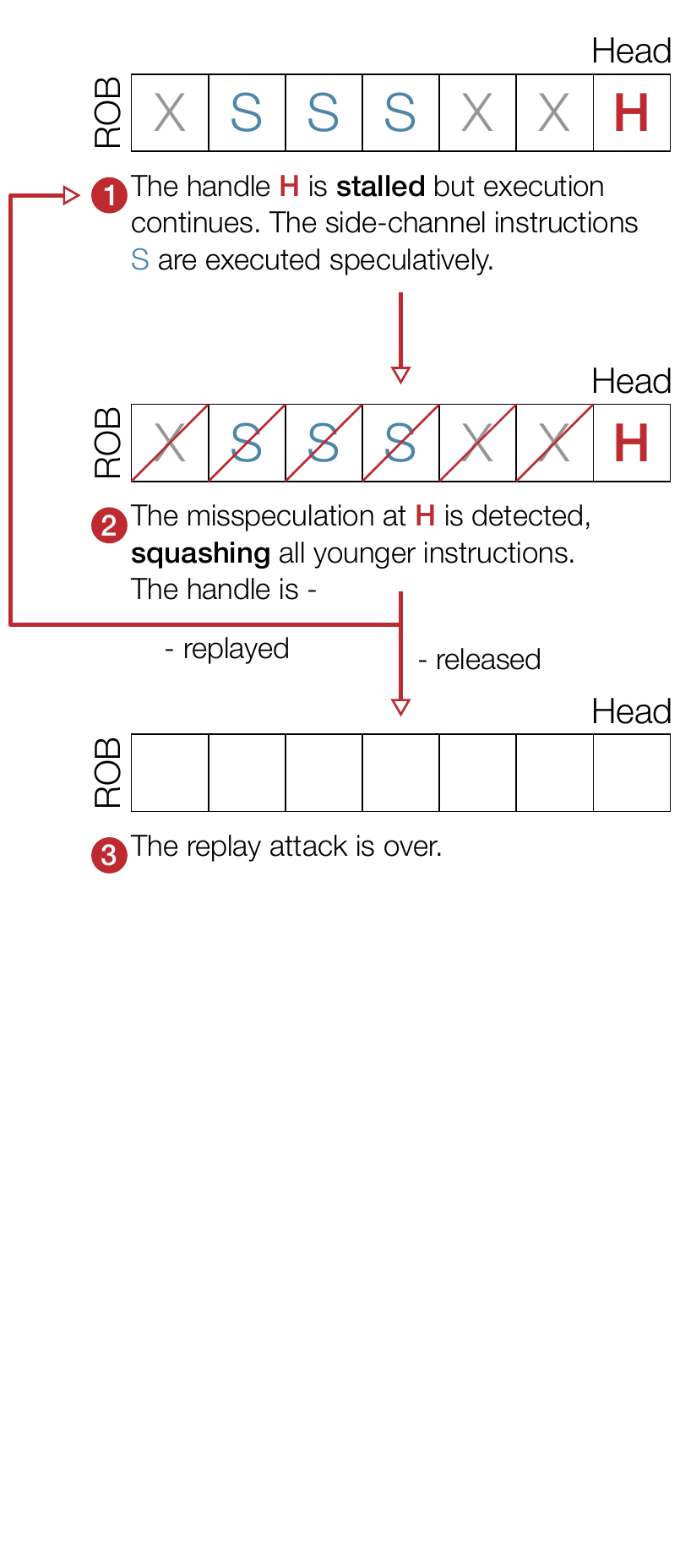}
    \caption{Single Handle}
    \label{fig:replay-attacks-1}
  \end{subfigure}\hfill%
  \begin{subfigure}[t]{0.2960\textwidth}
    \includegraphics[width=\textwidth]{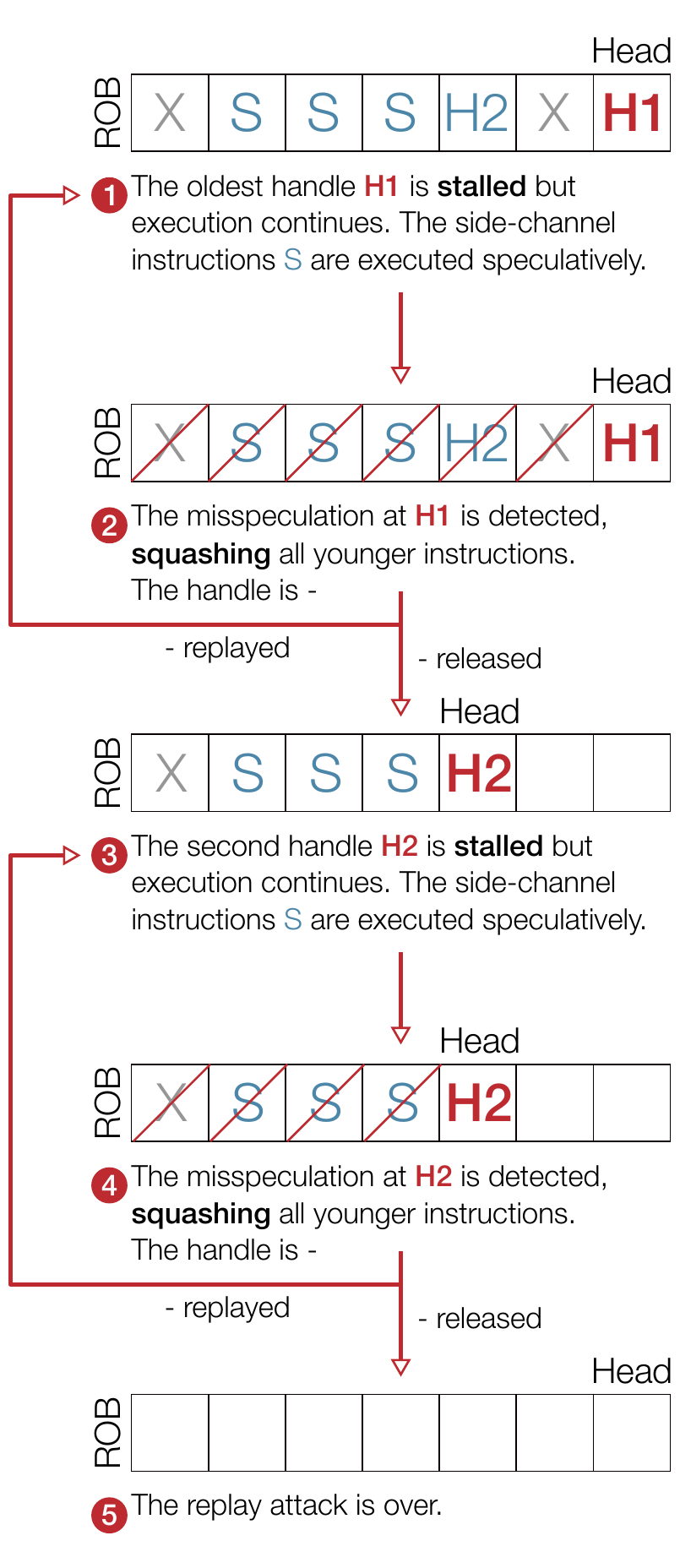}
    \caption{Serial Handles}
    \label{fig:replay-attacks-2}
  \end{subfigure}\hfill%
  \begin{subfigure}[t]{0.3105\textwidth}
    \includegraphics[width=\textwidth]{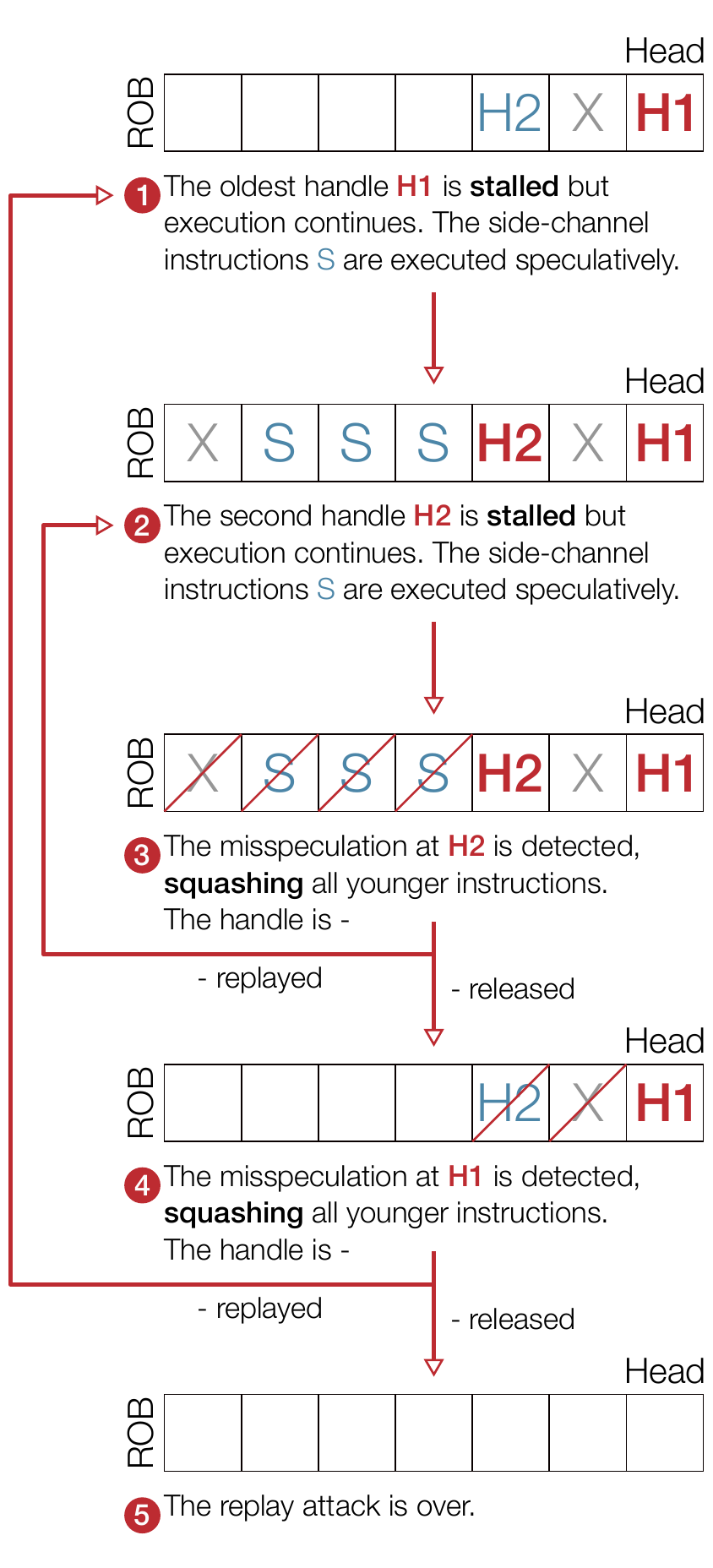}
    \caption{Nested Handles}
    \label{fig:replay-attacks-3}
  \end{subfigure}
  \caption{Three different microarchitectural replay attack patterns. The
  instructions in the reorder buffer are marked with an `H'~for the replay
  handles, `S' for the side-channel instructions, and `X' for other instructions
  not specific to the attack. The currently acquired handles are denoted in
  bold.
}
  \label{fig:replay-attacks}
\end{figure*}

Many modern CPUs offer secure execution contexts referred to as \emph{trusted
execution environments} (sometimes also referred to as ``secure enclaves'') that
protect the executed code from outside interference, including interference from
the operating system (OS) or the hypervisor. %
The characteristics of these enclaves differ for each architecture, but they
typically include encrypted memory for applications running in the enclave, code
verification to prevent malicious code from being executed in the enclave, and
hardware-enforced isolation of the enclaved execution context from any other
execution context in the system, including the OS. %
These measures are meant to protect sensitive code and data, such as
cryptographic functions and their keys, from any attacker that might have
compromised other parts of the system, including the OS or the hypervisor. %
MicroScope targets exactly this case, focusing specifically on Intel's Secure
Guard Extensions (SGX) enclave as a use case, although the underlying
exploitable concept is not limited to SGX. %
% Specifically,
MicroScope exploits the fact that under SGX the page management of the
application is still delegated to the OS\footnote{This is considered secure
because the memory accessed under SGX execution is cryptographically encrypted
and verified, preventing even the OS from accessing or manipulating it.}, to
capture the execution of the application and force the application to be
re-executed as many times as necessary for the side-channel attack to be
successful. %

Specifically, MicroScope takes advantage of how page faults are handled during
execution and of the %speculative
out-of-order capabilities of modern CPUs.  %
Assuming that the victim code is known and a compromised OS, MicroScope works as
follows:

\begin{enumerate}

    \item The OS manipulates the page table of the victim application to ensure
      that a load instruction will cause a page fault. %
      This instruction is referred to as \emph{a handle}, and the operation is
      referred to as \emph{``acquiring a handle.''} %

    \item When the victim application reaches the handle (\textbf{H} in
      \autoref{fig:replay-attacks-1}), a TLB miss occurs and a page walk
      is triggered. %
      The exact details of what happens during the page walk are beyond the
      scope of this work, but the two main points of interest are that
      \begin{enumerate*}[label=(\roman*)]

          \item it will take some time before the page fault is resolved and

          \item during this time the execution of the victim application will
            continue speculatively.

      \end{enumerate*}

    \item While executing speculatively, the victim application executes the
      instructions that follow the load, as long as they are \emph{not} dependent on
      the value of the load.%
      \footnote{Unlike speculative side-channel attacks, these instructions are
      slated to execute and commit. If fed with non-speculative data from before
      the handle \textbf{H}, they cannot even be delayed by proposed defences,
      e.g.,~\cite{weisse_nda:_2019,yu_speculative_2019,sdo20}.} %
      In the case of MicroScope, the handle is selected so that the instructions
      that are speculatively executed following the page fault unwillingly
      form the transmitter for the side-channel (\textbf{S}~in
      \autoref{fig:replay-attacks-1}, step \circled{1}). %
      % {\color{red} Unlike speculative side-channel attacks (e.g., Spectre), \textbf{S} instructions
      % %Note that \textbf{S} instructions, despite being temporarily speculative,
      % are slated to execute and commit.
      % Furthermore, if fed with non-speculative data from before the handle \textbf{H},
      % they cannot even be delayed by NDA~\cite{weisse_nda:_2019},
	  % STT,~\cite{yu_speculative_2019}, SDO~\cite{sdo20}, or similar techniques.}
      % Chris: not only we are criticizing NDA and STT way too often, I don't
      % think it's a good idea to go into this in the example. There aren't
      % enough details in the example to know what NDA or STT would do in this
      % case. This can open us up to criticism.

    \item While the victim is executing the side-channel instructions, the page
      walker will conclude that the page is not mapped and trigger a page fault,
      delegating the page handling to the OS. %
      This causes all of the instructions after the load, including the
      transmission code that has already been executed once speculatively, to be
      squashed and later re-executed %/replayed
      (step \circled{2} in \autoref{fig:replay-attacks-1}). %

    \item The OS updates the page table and invalidates the relevant TLB
      entries, signaling the victim application that the load is ready to be
      executed. %
      However, the OS does this in a way that %will
      ensures that the handle \emph{will page fault again}, and execution will
      restart from the second step. % listed above. %
      %\cs{This is step 2 from the list, not the figure, how to avoid confusion?}

    \item By repeating this procedure, the OS is able to keep the victim
      application in a speculative loop, where the handle will keep faulting,
      causing the instructions that follow it (the transmission code) to be
      speculative executed again and again. %
      The only way for the loop to be broken is for the attacker to
      \emph{release} the handle, allowing the victim to successfully service the
      TLB miss. %

\end{enumerate}

Essentially, by abusing the speculative execution mechanism of squashing and
re-executing, MicroScope is able to trap the execution of the victim application
in a loop for an arbitrary number of \emph{replay iterations}, until the
attacker is able to reliably denoise the side-channel. %
The receiver in this case can be a typical side-channel receiver, trying to
detect the side-effects caused by the transmitter, such as changes in the cache
or FU contention. %
% Since the attacker can control the execution of the application on a
% fine-grained level, it is possible to fully synchronize the transmitter with the
% receiver. %
We will not go into details about the receiver, as our solution focuses on
disrupting the transmitter; instead we point any curious readers to the original
MicroScope paper~\cite{skarlatos_microscope:_2019}. %

It is also possible to use more than one handle, as seen in
\autoref{fig:replay-attacks-2}. %
The attacker simply releases the first handle and then acquires a second handle,
assuming that more than one handle can be found in the %short
region of code between the first handle and the side-channel instructions. %
Under the conditions described in MicroScope using multiple handles in this way
is not useful, but it can be abused by yet unknown attacks. %

%\subsection{Microarchitectural Replay Attacks Beyond MicroScope}
\subsection{Replay Attacks Beyond MicroScope}
\label{sec:beyond-microscope}

MicroScope exploits page faults and a specific behavior found in some secure
enclaves (delegating page faults to the OS), but that does not mean that this is
the only behavior that can be exploited. %
Specifically, under MicroScope, a single instruction that page faults is used as
a handle to replay a set of instructions indefinitely
(\autoref{fig:replay-attacks-1}). %
This is possible because
\begin{enumerate*}[label=(\roman*)]
    \item page faults are a specific type of misspeculation that can be
      repeated indefinitely, and

    \item there is nothing in the architecture that prevents an instruction
      from misspeculating several times in a row.
\end{enumerate*}
However, as the authors of MicroScope allude to in their work, it is conceivable
that other forms of speculation can be used as handles, such as branch
prediction or even transactional memory\footnote{Transactional memory is not
necessarily implemented as conventional speculative execution, confined within
the ROB, but it does have similar characteristics.}, neither of which can be
repeated indefinitely. %
Once a branch has been executed, the correct path is known and the
%same instruction (dynamic instruction in the ROB)
incorrect path will not be misspredicted a second time. %
Similarly, transactions usually abort after a number of tries, at which point
they follow a fallback path. %
By using multiple handles (\textbf{H1}~and \textbf{H2} in
\autoref{fig:replay-attacks-2}), an attacker could extend the duration of the
attack for each case, especially if handles of different types are used
together. %
In addition, if the system restricted the number of times each instruction is
allowed to misspeculate and then be re-executed speculatively, e.g., as a simple
but na\"ive solution to MicroScope, an attacker would still be able to use
multiple handles to force a finite but tangible number of replays. %

We can see that multiple handles acquired and released one after the other can
be abused by an attacker to bypass some of the restrictions posed by different
kinds of speculation and system restrictions. %
However, using handles \emph{serially} like this only allows for a limited
number of replays, with the total number being the sum of the replays of
each handle. %
In contrast, \autoref{fig:replay-attacks-3} shows a way of \emph{nesting}
handles where each handle (\textbf{H1} in the figure) amplifies the number of
replays of the next handle (\textbf{H2}). %
This works by
\begin{enumerate*}[label=(\roman*)]

  \item having the inner handle \textbf{H2} cause as many replays as
    possible before releasing it (\circled{2} and \circled{3} in
    ~\autoref{fig:replay-attacks-3}),

  \item using the outer handle to squash and then re-execute the inner handle
    (\circled{1} and \circled{4}),

  \item re-acquiring the new (as far as dynamic instructions are concerned)
    inner handle, and

  \item repeating indefinitely. %

\end{enumerate*}
With this technique, the total number of replays is not the sum of each
individual handle's replays, but instead the product, growing exponentially
with each handle. %
This can make a huge difference in the number of replays. %
For example, if the attacker is able to acquire five handles and use each only
once (i.e., one replay each), with the serial handles the total number of attack
iterations would amount to ten, while with the nested handles it would amount to
32. %
Of course, nesting handles has its own challenges, including the fact that not
all forms of speculation can be used; in some cases (e.g., page faults) the
misspeculation is not handled until after the instruction has reached the head
of the ROB. %
On the other hand, modern microarchitectures allow branch prediction to go
several levels deep, resulting in the possibility of multiple outstanding
(unresolved) predictions at any one time. %
By arranging for \emph{older} branches to depend on \emph{longer latency}
operations (e.g., misses deeper in the memory hierarchy) and thus resolve slower
than younger branches, an effective nested multiple-handle attack can be
mounted. %
Similarly, if \emph{closed nesting} of transactions is
supported~\cite{haines1994composing}, the same effect can be easily achieved by
controlling the abort of the inner versus outer transactions. %
In closed nesting, an abort of the inner transaction does not abort the outer
transaction which will then proceed to repeat the inner transaction. %

Our goal is not to simply stop the currently known attacks (namely MicroScope)
but to secure speculative execution from as many present and future
microarchitectural replay attacks as possible. %
Hence, we assume that all of the three cases presented here are possible and we
have designed and evaluated our solution accordingly. %

\section{Threat Model}
\label{sec:threat-model}

MicroScope, and microarchitectural replay attacks in general, make sense only
under certain conditions. %
For example, in the specific case of MicroScope, the assumption is that the OS
is restricted from accessing the victim's execution state and memory. %
If this is not the case, then the OS can manipulate the victim directly, without
the need for MicroScope. %
At the same time, even if direct manipulation is not possible, there are cases
where it is possible to repeat the execution of the side-channel without using a
microarchitectural replay attack. %
For example, a digital rights management (DRM) system that accepts encrypted
data and decrypts them with a secret key could be fed the same data repeatedly
by a malicious attacker, allowing for enough iterations to reliably leak the
secret key. %
This is possible because, in this case, the secret is immutable, regardless of
the input data. %
On the other hand, as a counter example, there exist applications where the
secrets consist of transient data, such as SGX implementations of
Tor~\cite{kim_sgx-tor_2018} (where the sensitive data are arbitrary network
packets) or applications where countermeasures against rollback attacks have
been taken~\cite{matetic+:SECURITY2017rote}. %
In such cases, the attacker only has a single opportunity to successfully leak
the data. %
These are exactly the cases where MicroScope and microarchitectural replay
attacks in general become useful. %
Coincidentally, these are also the cases we expect to be the least protected
against side-channel attacks, as, before MicroScope, they were very hard to
exploit. %

While MicroScope focuses on secure enclaves, one can imagine how a
microarchitectural replay attack can be performed from a non-privileged context
as well. %
For example, if the branch predictor is abused to create handles, it should be
possible to do a microarchitectural replay attack from an unprivileged SMT
thread. %
We observe that for a microarchitectural replay attack to be necessary, two
conditions have to apply:

\begin{enumerate}

\item The side-channel used in the attack is ineffective unless it is repeated
  numerous times, as otherwise there would be no reason to replay the attack
    code. %

\item The attacker cannot arbitrarily execute the victim code. %
  If this was not the case then there would be no need for a microarchitectural
    replay attack as the attacker could simply re-execute the victim code as
    many times as necessary. %

\end{enumerate}

These two conditions combined mean that it is completely safe, under our current
threat model, to allow a single execution of a side-channel attack while under
speculation, as it will not be effective. %
Any additional attempts will be blocked by our replay-blocking defence
mechanism, Delay-on-Squash. %

\section{Delay-on-Squash}
\label{sec:dos}

The principle behind Delay-on-Squash is simple: If an instruction is issued,
then squashed due to a replay handle, and then re-appears in the pipeline, it
is not allowed to be re-issued, as it might constitute part of a
microarchitectural replay attack. %
In the next part of this section, we have split the description of the
Delay-on-Squash mechanism into two subsections: %
First, we discuss the concept on a high level, with no regard for any
implementation constraints (\autoref{sec:dos-concept}), and then we discuss
the practicalities of an actual implementation
(\autoref{sec:dos-implementation}). %

% {\color{red} STEFANOS: This is very important but not explained later on.
% I don't see where you explain it in the implementation. I would like to see it clearly explained later on.}
% To provide an absolute defence against microarchitectural replay attacks,
% without introducing large performance overheads, we need to differentiate
% between code and replay iterations, leaving the former unhindered while
% disallowing the latter. %
% Chris: I will remove this for now, as it will be replaced by an explanation
% regarding loops and false positives in the perfect filter.

\subsection{Conceptual Description}
\label{sec:dos-concept}

To keep track of instructions that have been issued, squashed, and re-issued
under the same handle, Delay-on-Squash needs to keep track of all the potential
handles in the ROB. %
Due to the serial and nested cases that we described in
\autoref{sec:beyond-microscope}, it is not enough to keep track of only the
handle that caused the last squash, instead Delay-on-Squash %needs to
takes into consideration all the handles that affect each squashed
instruction. %
To prevent interference from other contexts, this information needs to
accessible only by the enclave and should be securely stored and restored on a
context switch, much like the rest of the execution state (e.g., registers). %

% To achieve this,
To track all the handles, Delay-on-Squash utilizes a FIFO queue where all the dynamic
instructions that can cause misspeculation and squashing are inserted during
dispatch. %
While in the queue, these instructions are considered as potential handles and,
by definition, are considered as ``unsafe.'' %
Instructions can only be removed from the head of the queue\footnote{Even if the
instructions are squashed in the ROB, they still need to remain in the handle
queue until they reach the head.} and only if it can be determined that they are
``safe,'' i.e., that they can no longer act as a handle, which happens only when
they have \emph{moved outside the window of speculation}. %
We will discuss the exact conditions for this in \autoref{sec:dos-shadows}, but
for now let us say that this happens when the instructions are ready to be
committed. %
As the queue is a FIFO, and handles can only be removed from the head of the
queue, for a handle to become safe all the older handles %(in ROB order)
in the queue need to also become safe. %
This is a very important condition, as it prevents the serial and nested replay
patterns presented in \autoref{sec:beyond-microscope}. %

With this queue of potential handles, it is now possible to keep
track of squashed instructions on a per-handle basis. %
Specifically, every time a misspeculation occurs and instructions are squashed,
Delay-on-Squash records the PCs for all the instructions that have been issued
and are now being squashed. %
The \emph{youngest} handle (in ROB order) is retrieved from the queue and is
%stored alongside the squashed PCs. %
associated to the squashed PCs. %
These PCs then remain stored until their corresponding handle is determined to
be safe, i.e., when the handle is removed from the queue. %
Using the youngest handle, instead of the actual misspeculating instruction that
caused the squash, is important, once again to prevent the nested replay
pattern. %
Essentially, by storing the PCs of the squashed instructions until the youngest
(at the point of the squash) handle is safe, Delay-on-Squash ensures that the
record of the squashed PCs will remain stored until all the handles that were
present in the window of speculation during the squash have left the window and
are thus safe. %
With this guarantee, all that remains is to check these records before issuing
an instruction. %
If an instruction PC matches one in the stored records, it means that this
instruction has previously been issued and squashed, and that \emph{the handles
that preceded it are still in the ROB and are still considered unsafe}. %
Such instructions are prevented from being issued until the relevant handles are
deemed to be safe and the records are removed. %

With this mechanism in place, we can instantly detect when an instruction has
been issued, squashed, and then re-issued, even in more complex cases, such as
when the attacker might be utilizing nested handles. %
This enables us to detect and prevent replay iterations of the code, instantly
stopping any microarchitectural replay attacks in their tracks. %
This mechanism does have one disadvantage: %
The pattern of ``issue, squash, and re-issue'' can also happen under normal
speculative execution (i.e., when not under attack), for example when a load is
squashed due to a memory order violation, as in such cases the execution path
remains the same. %
In addition, if we are executing a loop that is small enough to fit several loop
iterations in the window of speculation, a squash in one of them will cause all
the iterations that follow (within the window) to be delayed, as the
instructions at each iteration all share the same PC. %
% \cs{Talk about loops instead of loads here (discussion with Stefanos).}
Fortunately, as we will see in the evaluation (\autoref{sec:evaluation}), the
actual number of cases affected by this are very small and come at a negligible
performance cost ($1\%$). %

% \vspace{\baselineskip}
% \hms{The following paragraph can be removed to save space.}
% In the next section we will discuss the actual implementation of
% Delay-on-Squash, taking into consideration the area, latency, and energy costs
% of the mechanism. %
% We will also provide a step-by-step example of how the mechanism works, which we
% have omitted from the conceptual description due to space constraints. %

% Depending on the reason for the squash, it can be less or more likely than
% an instruction is seen again after the squash.
%   For example, during a branch missprediction, the instructions that follow the
% branch should be mostly different after the squash, as a new path is taken, with
% the exception of instructions that follow after the reconvergence point of the
% branch. On the other hand, if the squash is caused by a load that has violated
% the memory order, then the instructions that follow can be exactly the same as
% before.

%\subsection{In Practice}
\subsection{Efficient Implementation}
\label{sec:dos-implementation}

While the Delay-on-Squash mechanism, as conceptually described in the previous
section, can be implemented into an actual hardware solution, it would require
large storage and expensive content addressable memories for keeping
\emph{exact sets} of squashed PCs. %
This would lead to prohibitively large area, latency, and energy overheads. %
Instead, we use Bloom filters~\cite{10.1145/362686.362692} to represent the sets
of PCs (of squashed instructions) that are temporarily prevented from being
issued speculatively. %
With this approach, we can store the PCs of tens of squashed instructions with
only a few bits of storage. %

% {\color{red}
% In this section, we will describe a more practical implementation, which
% sacrifices a tiny amount of performance to false positives ($1-2\%$ below the
% conceptual mechanism), but dramatically reduces the hardware overheads. %
% Specifically, we will describe how we can avoid storing all the squashed and the
% individual handles by utilizing a number of rolling Bloom filters
% (\autoref{sec:dos-bloom}). %
% In addition, we will also discuss how to determine when a handle becomes safe as
% early as possible, and how to efficiently track the speculative status of the
% instructions in the ROB (\autoref{sec:dos-shadows}). %
% }

\subsubsection{Bloom Filters}
\label{sec:dos-bloom}

Bloom filters are hash-based, probabilistic data structures used to test if an
element (in our case, a PC) is part of a set. %
They work by hashing the element with a number of hash functions, each of which
indicates a position in the filter that has to be set to ``one''. %
While it is possible to get a false positive when checking the filter, it is not
possible to get a false negative, which is an important property for us, as
false negatives would lead to unsafe replay iterations. %
False positives, on the other hand, only manifest as reduced opportunities for
speculation, in our case causing a negligible performance overhead, when
compared against the overheads of existing security mechanisms in secure
enclaves~\cite{zhao16}, which are the primary target for Delay-on-Squash. %

\fakeheader{Bloom Filter Implementation}
In Delay-on-Squash we use the simplest, most efficient form of binary Bloom
filters where the only way of erasing elements from the filter is to clear the
whole filter by bulk-resetting it. %
Other approaches also exist~\cite{cohen2003spectral, fan2000summary}, but they come at increased overheads and
they would not offer significant performance benefits for our use case. %
For our implementation, we assume that all the hash functions of the PC of a
particular dynamic instruction are precomputed during dispatch and kept in its
ROB entry. %
The precomputed hash functions are then used (in the case of a squash) to index
the Bloom filter and set the corresponding ``ones,'' in parallel with multiple
other ROB entries. %
We also assume that the whole process is hidden behind the back-end recovery
latency following a squash~\cite{gonzalez2010processor}. %

% CHRIS: I am leaving this here as comments, but since we are already over the
% page limit I think we should skip the explanation why using complicated Bloom
% filters is a problem.

% \textbf{Why Binary Bloom filters?}
% In addition to false positives,
% Bloom filters have another property that we need
% to account for when designing the Delay-on-Squash implementation:
% Removing individual elements (PCs) from a Bloom filter is expensive.
% To do that, we need to resort to \emph{counting} Bloom filters~\cite{Bloom} (or
% one of their many varieties), significantly increasing our cost.
% Moreover, updating a counting Bloom filter entails several (as many as the hash
% functions) \emph{read-increment-write} operations per PC.
% Parallelizing the insertion of multiple PCs in the Bloom filter becomes
% problematic as we have to watch out for conflicts and serialize conflicting PCs.
% That can be too expensive to do (either in cycles or in read/write ports) when
% we need to store tens of PCs on a squash.
% In contrast, a simple binary Bloom filter can be updated much faster by just
% setting all the new ``ones'' in place.

\fakeheader{Rolling Bloom Filters}
In the conceptual description, for each squash, we associate each set of
squashed PCs with a handle. %
Observe, however, that in the case of a replay attack, the set of PCs would
hardly change from squash to squash. %
Maintaining the same \emph{redundant} information across multiple Bloom filters
is, clearly, a waste of resources. %
Instead, we could consolidate the information in a single rolling Bloom filter
spanning several squashes, holding all the PCs of all the squashed instructions.
In each squash, we insert the squashed PCs in the filter and then we
\emph{associate the Bloom filter with the youngest handle in existence in the
handle queue.} %
While this is an effective way of keeping all the information we need to prevent
replay attacks, it makes the clearing of the Bloom filter difficult. %
Recall that we have opted for a simple binary Bloom filter, where it is not
possible to erase individual instructions. %
% ALTERNATIVE TEXT BELOW
% In addition, as we only store the youngest handle across all squashes, even if
% we could erase individual instructions from the filter, we would not know which
% instructions can be safely erased at any point. %
% \hms{We also don't have the original PCs (hashes) so even with the
%   above to there is not enough information to clear individual
%   PCs. So is the information that individual PCs can't be cleared
%   something that has to be discussed or simply focus on the simplicity
% of storing the youngest handle.}
% You can now see the problem: the condition for bulk-resetting the Bloom filter
% %,
% %which is the only way to erase instructions from the filter,
% is for the
% associated handle to leave the window of speculation and become safe. %
% %CHRIS: I removed the footnote here, as we already say that we will explain this
% %later. I just need to add something about "windows of speculation" in the
% %shadows section.
% Unfortunately, as at each squash we re-associate the Bloom filter with the
% \emph{youngest} handle during the squash, the lifetime of the filter might be
% extended an arbitrary number of times and the clearing might be deferred for an
% arbitrary long time. %
% NEW TEXT
Also, we only associate the youngest seen handle with the Bloom filter, since it
is not possible to remove individual PCs anyway. %
The condition for clearing the Bloom filter is then for the associated handle to
leave the window of speculation and become safe, i.e., all PCs in the Bloom
filter are safe. %
Unfortunately, as at each squash we re-associate the Bloom filter with the
\emph{youngest} handle during the squash, the lifetime of the filter can be
extended an arbitrary number of times and the clearing can be deferred for an
arbitrarily long time. %

\begin{figure}[t]
  \centering
  \includegraphics[width=0.65\columnwidth]{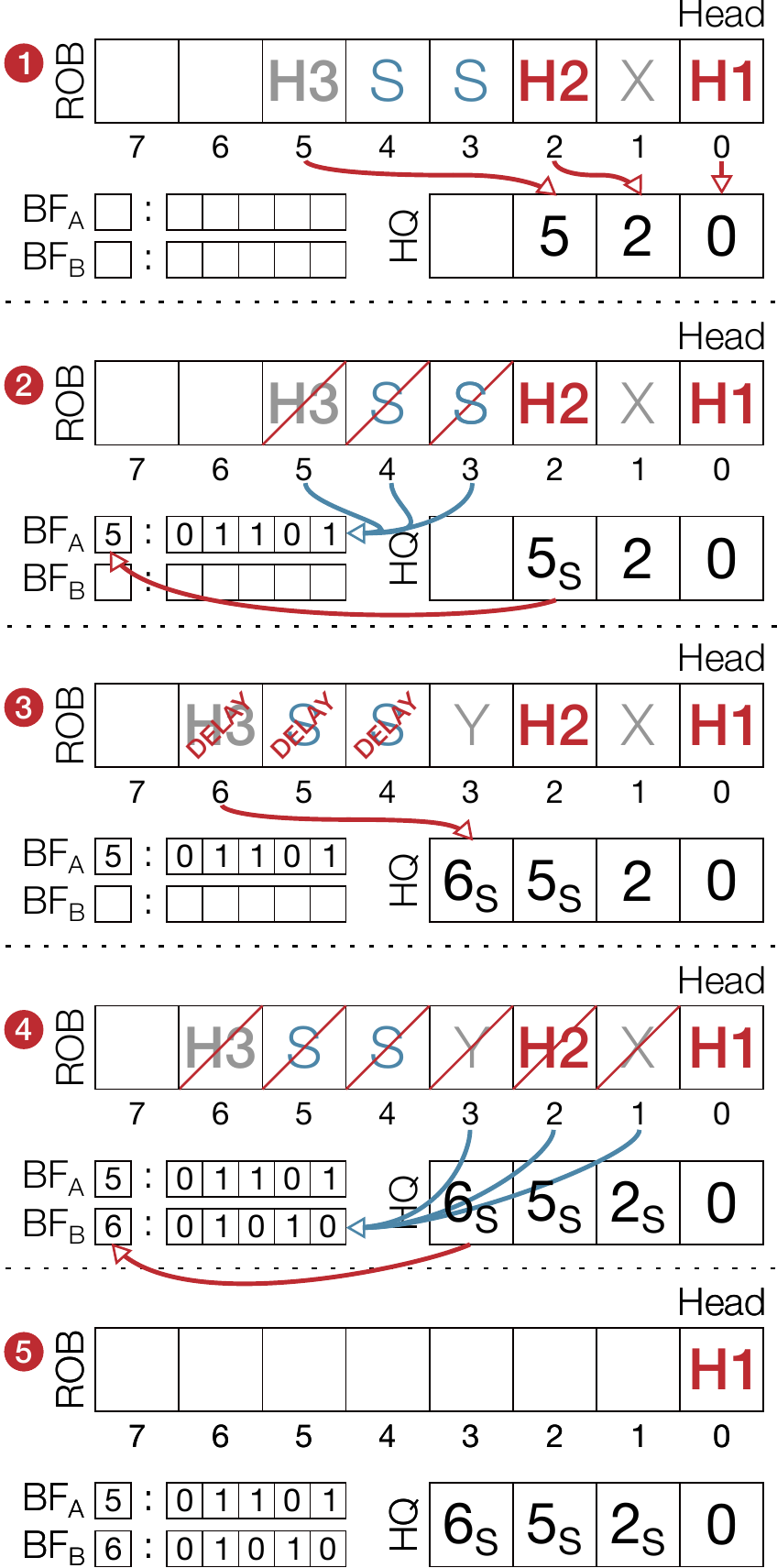}
  \caption{Step-by-step example of the instruction tracking mechanism in
  Delay-on-Squash.}
  \label{fig:example}
\end{figure}

Instead, we use two rolling Bloom filters and switch between them
periodically. %
At each point in time, one of the filters is designated as \emph{active}, while
the other is inactive, \emph{waiting to be cleared.} %
On a squash, the PCs of the squashed instructions are inserted in the currently
active filter, and the filter is associated with the youngest handle. %
Meanwhile, the inactive filter is waiting for its associated handle to leave the
window of speculation in order to be cleared with a bulk-reset. %
As soon as the inactive filter is cleared, then it can take over the role of the
active filter, letting the previous active filter become inactive and eventually
be cleared as well. %
In our implementation of Delay-on-Squash, we have also included a mechanism that
checks if a filter is starting to become saturated (based on the number of bits
that are set to ``one'') before switching to a new active filter, to better
space out the filters over the window of speculation. %
% An important detail here is that during this process,
Of importance here is that
instructions that are about to be issued need to check both filters, the active
and the inactive, to see if they can be issued safely. %
The approach can naturally be extended to a cyclical list of several Bloom
filters, out of which one is active and the rest are inactive. %
% In the evaluation section, we will see how the number of filters, in conjunction
% with the number of bits per filter, affects the performance of the system. %

The Bloom filters (as well as the shadow tracking information) are
context-specific, i.e., each execution context has its own set of filters, which
is securely stored and reloaded on a context switch. %
This prevent other contexts, including the OS, from saturating or otherwise
manipulating Delay-on-Squash. %

\fakeheader{A Working Example}
Figure~\ref{fig:example} contains a step-by-step example of how Delay-on-Squash
tracks handles and how the Bloom filters ($BF_A$ and $BF_B$ in the
figure) are used: %

\begin{enumerate}

    \item The ROB contains three potential handle instructions, $H1$, $H2$, and
      $H3$. %
      % \hms{Maybe we should just rename it to H3 to avoid having to talk about
      % oracle knowledge.} %
      % In this example, we have oracle knowledge that only $H1$ and $H2$ are
      % actually used as handles, so they are marked differently than $H?$. %
      The example also contains the instructions $X$ and $S$, with $S$ being
      actual side-channel instructions, and $X$ being some other
      instruction. %
      The exact types of all these instructions are not important for this
      example. %
      In the first step, we see how all the potential handle instructions are
      inserted into the handle queue (HQ).

    \item In the second step, we detect that $H2$ was misspeculated and needs to
      be re-executed. %
      For simplicity, we assume that the instruction that caused the squash
      remains in the ROB. %
      All the younger instructions that follow it (and have already been issued)
      are hashed and inserted into the active Bloom filter ($BF_A$). %
      The filter is then marked with the youngest potential handle, which can be
      found at the tail of the handle queue. %
      Handles can only be removed from the handle queue when they reach the
      head so all handles that are squashed are marked accordingly (5$_S$). %

    \item Execution restarts from $H2$, this time following a slightly different
      path, issuing all the same instructions as before, as well as the
      additional instruction $Y$. %
      For the example, we assume that $H2$ is still not considered safe. %
      $S$ and $H3$ hit in $BF_A$ and are delayed, while $Y$ does not hit and is
      allowed to execute speculatively. %
      We have now successfully prevented the side-channel instructions $S$ from
      being replayed, with the unwanted side-effect of also delaying the
      innocuous $H3$ instruction. %
      However, we will see in the evaluation section that delaying some
      additional instructions introduces only negligible performance overheads.

    \item Now we detect that $H1$ was misspeculated and all younger instructions
      need to be squashed. %
      $X$, $H2$, and $Y$ are the only younger instructions that were actually
      issued (since the rest were delayed), so we insert them in the Bloom
      filter. %
      Since $BF_A$ is already more than half full, we instead update the $BF_B$
      filter, as seen in the figure. %

    \item Finally, in the last step, we see that all handles, except $H1$ have
      been marked as squashed. %
      However, since $H1$ remains, we cannot reset any of the filters, as they
      are marked with instructions younger than $H1$. %
      Because of this, $H1$ can cause an infinite number of squashes (e.g., by
      going back to the ROB state in step \circled{1} in the example) and
      repetitions, but the side-channel instructions $S$ will not be replayed
      again. %
    \item Once the $H1$ handle is resolved it is removed from the HQ upon which
      the squashed handles reaches the head and can trigger the Bloom filters to
      be cleared (not shown). %

\end{enumerate}

In some rare cases, it is possible that after squashing the handle queue is left
empty, as all instructions are either squashed or deemed safe (e.g., if $H_1$ is
safe). %
In such cases we can run into a corner case where the squashed handles ($H_2$)
are re-introduced into the window of speculation but the Bloom filters have all
been cleared, enabling the attacker to perform several replay iterations. %
We handle this conservatively by delaying the clearing of the Bloom filters by
the length of the dynamic instruction window, to ensure that the handles are not
re-introduced. %
This happens very rarely and has no measurable effect on performance. %

% It should be noted that while in the example we assume oracle knowledge of
% which instructions are used as handles and which are used as side-channels the
% Delay-on-Squash mechanism does not rely on such information. %
% We make the distinction only to make the example more clear. %

% \subsubsection{Speculative Instructions and Shadows}
\subsubsection{Handles and the Window of Speculation}
\label{sec:dos-shadows}

We have talked about handles that can cause misspeculation and squashing and
when such handles can be considered as safe. %
In the most na\"ive approach, we can consider handles to be safe when they reach
the head of the ROB and are retired, but this is awfully pessimistic. %
Instead, we draw from the existing research on speculative
execution~\cite{mehdi_ooo, yan_invisispec:MICRO2018, ghosts, dom19, dom20,
invarspec20, bell_lipasti:ISPASS2004} and consider handles as safe when they can
no longer cause squashing, regardless of their position in the ROB. %
Specifically, we have adopted the approach of using speculative shadows by
Sakalis et al.~\cite{ghosts, dom19, dom20}, a mechanism for detecting the
earliest point at which an instruction is no longer speculative. %
While these shadows are designed to work with speculative side-channel defences,
which do not necessarily work against microarchitectural replay attacks
(\autoref{sec:related-work}), the underlying principle can still be used. %

According to Sakalis et al., any speculative instruction that can cause
squashing is referred to as a ``shadow-casting'' instruction. %
Depending on the type of speculation, Sakalis et al. have defined four different
types of shadows~\cite{ghosts, dom19}, but these can be extended to include
other types of speculation as well, such as the transactional memory case
described earlier. %

\begin{itemize}

  \item \textbf{E-Shadows} are cast by exceptions, as is the case of the page
    faults used in MicroScope. %

  \item \textbf{C-Shadows} are cast by control flow instructions such as
    branches. %

  \item \textbf{D-Shadows} are cast by stores with unknown addresses, as they
    might modify values of speculatively executed~loads. %

  \item \textbf{M-Shadows} are cast by speculative reordering that might violate
    the memory model of the system. %

\end{itemize}

Once an instruction
\begin{enumerate*}[label=(\roman*)]
    \item is no longer shadowed by another instruction and

    \item no longer casts any shadows itself,
\end{enumerate*}
i.e., when there is no reason for said instruction to be squashed, the
instruction is considered non-speculative. %
At this point, the instruction has left the window of speculation and, assuming
that the instruction was a potential handle, it can be considered as safe. %
The advantage of this approach is that it is possible for potential handles to
reach the safe state a lot earlier than if we were waiting for them to retire. %
In addition, Sakalis et al. also describe a hardware implementation to track
speculation based on a FIFO queue~\cite{dom19}, where younger shadow-casting
instructions are only resolved once all older shadow-casting instructions have
also been resolved. %
This design fits well with Delay-on-Squash, as the handle queue has similar
characteristics. %

Note that while using the speculative shadows is not necessary and the use of
alternative methods is possible (e.g., using the head and tail of the ROB),
doing so in a na\"ive manner can lead to significant performance
degradation~\cite{ghosts, invarspec20}. %

\subsubsection{Security Implications}

As discussed in (\autoref{sec:threat-model}), Delay-on-Squash only targets
microarchitectural replay attacks, where more than one iteration is necessary to
leak sensitive information. %
Sometimes, it is possible to leak information with a single iteration of an
attack, although this is not easy to achieve~\cite{lyu_survey_2018,
aldaya_port_2019, maurice_hello_2017}, especially when the attacker does not
have full control of the victim. %
Delay-on-Squash does not protect against such attacks, as they are
indistinguishable (and are, in fact, part of) normal execution. %
Instead, if such protections are required, Delay-on-Squash can be combined with
other defence mechanisms, such as defences against speculative side-channel
attacks. %
We direct the reader to out Related Work (\autoref{sec:related-work}) for more
details. %

In addition, we want to ensure that Delay-on-Squash does not introduce any new
side-channels into the system. %
As already discussed, the state kept by the mechanism, i.e., the bloom filters
and the handle tracking information, need to be kept isolated from other
contexts and stored/restored on a context switch. %
This prevents the attacker from manipulating Delay-on-Squash either to ``make it
forget`` a replayed instruction or to introduce unnecessary overheads in the
victim application. %
At the same time, the attacker cannot in any way probe the information that the
Delay-on-Squash mechanism has of the victim, as that would allow the attacker to
ascertain which instructions the victim has executed. %
These can be enforced by the hardware by isolating the mechanism between
contexts and storing it in an encrypted manner on a context switch, much like
the rest of the context (e.g., the register file) is already stored. %

Finally, as Delay-on-Squash prevents instructions from executing under certain
conditions, it raises the question of whether it can be used to mount a
Denial-of-Service attack on the victim. %
While Delay-on-Squash can affect the performance of the victim negatively, as
long as instructions are being committed from the head of the reorder buffer
(which is never delayed by Delay-on-Squash) then execution will not stall. %
Execution can only stall if the victim is caught into an execute-squash-replay
loop (due to a microarchitectural replay attack), in which case no forward
progress would have been made even if Delay-on-Squash was not present. %
In practice, when execution is slowed down for any significant amount of time,
due to an attack, the hardware should notify the enclaved software, allowing it
to take appropriate measures to prevent the leakage and ensure that the system
does not stall. %

\section{Evaluation}
\label{sec:evaluation}

\begin{table}
  \centering
  \caption{The simulated system parameters.}
  \label{tbl:params}
  \begin{tabular}{ l | l }
    \hline
    Parameter & Value\\
    \hline
    Technology node                           & 22nm @ 3.4GHz \\
    Issue / Execute / Commit width            & 8 \\
    Cache line size                           & 64 bytes \\
    L1 private cache size                     & 32KiB, 8-way \\
    L1 access latency                         & 2 cycles \\
    L2 shared cache size                      & 1MiB, 16-way \\
    L2 access latency                         & 20 cycles \\
    \hline
    Number of Bloom filters                   & 2 \\
    Number of bits, hash functions per filter & 64, 2 \\
    \hline
  \end{tabular}
\end{table}

\begin{figure*}[t]
  \centering
  \includegraphics[width=\textwidth]{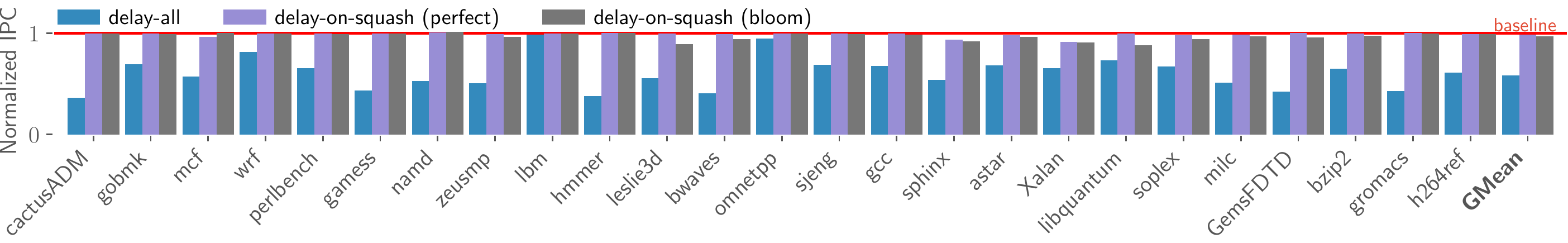}
  \caption{Performance (instructions per cycle) normalized to the baseline.}
  \label{fig:ipc}
\end{figure*}

% \begin{figure}[t]
%   \centering
%   \includegraphics[width=\columnwidth]{figs/squashes.pdf}
%   \caption{The total number of squashes due to misspeculation during the
%   execution of the benchmark. Note that the scale for the $y$-axis is
%   $\times10^{8}$.}
%   \label{fig:squashes}
% \end{figure}

\begin{figure}[t]
%             delay-replayed (bloom)
% Benchmark
% cactusADM                 0.000027
% gobmk                     0.005234
% mcf                       0.031949
% wrf                       0.001761
% perlbench                 0.003578
% gamess                    0.006544
% namd                      0.005028
% zeusmp                    0.006377
% lbm                       0.000047
% hmmer                     0.011141
% leslie3d                  0.020495
% bwaves                    0.007364
% omnetpp                   0.018879
% sjeng                     0.005611
% gcc                       0.008883
% sphinx                    0.008815
% astar                     0.009928
% Xalan                     0.016913
% libquantum                0.010976
% soplex                    0.015493
% milc                      0.019648
% GemsFDTD                  0.015421
% bzip2                     0.011326
% gromacs                   0.003288
% h264ref                   0.007677
% GMean                     0.005745
  \centering
  \includegraphics[width=\columnwidth]{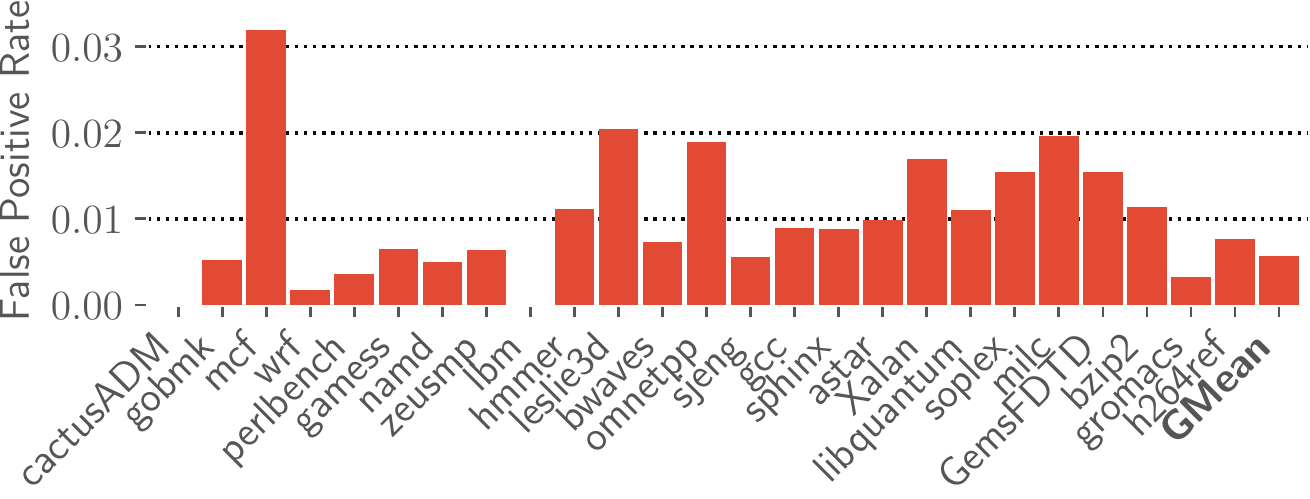}
  \caption{Rate of Bloom filter false positives over the total number of
  \emph{executed} (i.e., including squashed) instructions.}
  \label{fig:false-rate}
\end{figure}

We use the Gem5 simulator~\cite{binkert_gem5:CANEWS2011} together with the SPEC
CPU2006 benchmark suite~\cite{spec:cpu06} for the performance evaluation,
combined with McPAT~\cite{li_mcpat:MICRO2009} and
CACTI~\cite{li_CACTI:ICCAD2011} for the energy evaluation. %
% In Gem5,
We skip the first three billion instructions and then simulate in detail
for another three billion. %
The main parameters of the simulated system can be seen in
~\autoref{tbl:params}. %
To avoid mixing different energy models, we estimate the energy expenditure of
the additional handle and squash tracking structures as existing similar
structures in McPAT, appropriately sized. %
Specifically, the handle/shadow tracking mechanism is modelled as an additional
ROB structure, the Bloom filters as register files, and the hash functions for
the Bloom filters as an additional integer arithmetic unit. %
We evaluate the following configurations in detail: %

\begin{itemize}

  \item \textbf{baseline:} The insecure out-of-order CPU in Gem5. %

  \item \textbf{Delay-All:} A lower-bound, non-speculative configuration where
    all instructions are delayed until no unsafe handles precede them. %
    To make the comparison with the Delay-on-Squash configurations fair,
    Delay-All utilizes the same handle-tracking mechanism as Delay-on-Squash to
    detect when a handle is safe (\autoref{sec:dos-shadows}). %

  \item \textbf{Delay-on-Squash (Perfect):} The ideal implementation of
    Delay-on-Squash, as described in \autoref{sec:dos-concept}. %
    We assume that we have unlimited space for storing all the squashed
    instructions for as long as necessary. %

  \item \textbf{Delay-on-Squash (Bloom):} The actual implementation of
    Delay-on-Squash with Bloom filters, as described in
    \autoref{sec:dos-implementation}, with the Bloom filter parameters seen in
    \autoref{tbl:params}. %
    We chose these parameters empirically to balance performance, area, and
    energy usage. %

\end{itemize}

As the SPEC2006 applications do not contain any enclaved code regions, we apply
Delay-on-Squash during the whole execution of the application. %
In a real use case, Delay-on-Squash will only be applied to the regions of the
application running in the enclave, while regions (and applications) running
outside the enclave (as is the majority of applications on modern systems), will
not be affected. %

\subsection{Performance}
\label{sec:eval-performance}

\autoref{fig:ipc} contains the number of instructions per cycle (IPC),
normalized to the insecure out-of-order baseline. %
Overall, implementing Delay-All, which we are only including as a lower-bound
configuration, would lead to $41\%$ lower performance, when compared to the
baseline. %
On the other hand, implementing Delay-on-Squash with a perfect filter would
eliminate almost all of the performance cost, reaching $99\%$ of the baseline
performance. %
Only three applications see any perceptible performance degradation,
\texttt{Xalan} ($-8\%$), \texttt{sphinx} ($-6\%$), and \texttt{mcf} ($-3\%$),
all of which misspeculate and squash more often than average. %
In some cases, the same static instructions exist in the ROB in multiple dynamic
instances, for example when a loop is dynamically unrolled in the ROB
(\autoref{sec:dos-concept}). %
If a squash happens in such a situation, all instances of the static instructions
will be conservatively treated by Delay-on-Squash as potential replay attacks
and speculation will be restricted in a large part of the dynamic instruction
window. %
In contrast, applications like \texttt{cactusADM} and \texttt{lbm}, which rarely
misspeculate, are not affected at all. %
However, the amount of misspeculation in the application is not the only
parameter that affects its performance with the perfect filter, instead the
applications sensitivity to instruction and memory level parallelism (ILP and
MLP) is also important. %
Restricting the speculative execution of an application which relies on ILP and
MLP to achieve high performance can lead to more severe performance penalties,
when compared with an application that does not rely on either~\cite{dom20}. %

Finally, we have the realistic Delay-on-Squash implementation, based on the
Bloom filters, with the parameters described in \autoref{tbl:params}. %
Alongside the IPC, in \autoref{fig:false-rate} we can also see the false
positive rate of the applications, i.e., the number of false positives in the
Bloom filter over the number of instructions executed. %
On average, we observe one false positive every 174 instructions (a rate of
roughly $0.6\%$), but this number differs significantly between all the
application. %
Applications that have higher than average false positive rates (most
pronounced in: \texttt{mcf}, \texttt{leslie3d}, and \texttt{milc})
also tend to have lower performance than the baseline. %
However, once more, this false positive rate is not the only parameter affecting
the performance. %
The application whose performance is affected the most is \texttt{libquantum}
($-12\%$), which is not one of the applications with the highest false positive
rates. %
Instead, \texttt{libquantum} is a streaming benchmark that relies heavily on
MLP, so it is heavily affected by the false positives. %
In contrast, \texttt{omnetpp}, which is the application with the fourth largest
false positive rate, sees no performance degradation at all. %

Overall, for Delay-on-Squash with the Bloom filters, we observe a mean
performance of $97\%$ of the baseline, two percentage points lower than the
ideal version. %
Given that, by choice, applications running in secure enclaves exchange
performance for higher security guarantees, a $3\%$ performance degradation is
practically negligible. %

\subsection{Energy Usage}
\label{sec:eval-energy}

\begin{figure*}[t]

  \centering
  \includegraphics[width=\textwidth]{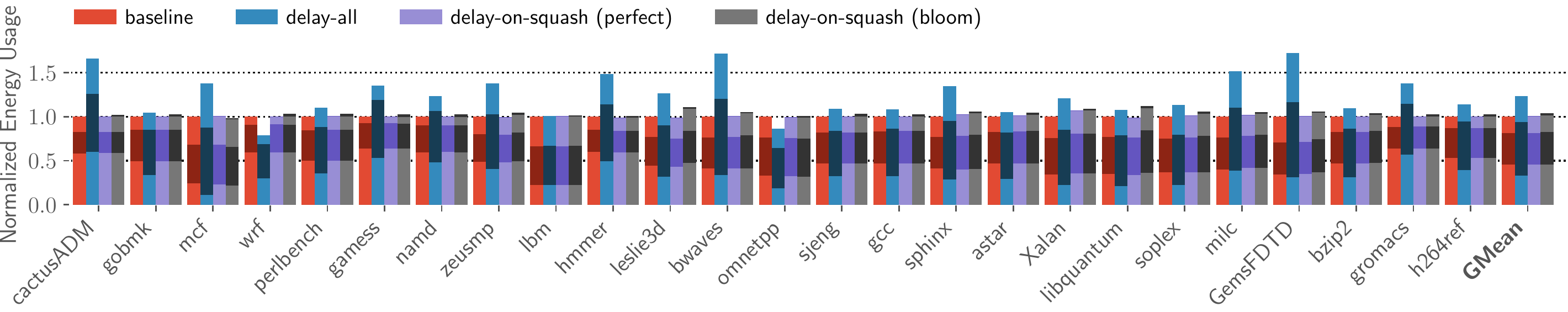}
  \caption{Energy usage normalized to the baseline. Each bar considers of four
  different parts, representing the dynamic CPU energy usage (bottom, light),
  the static CPU energy usage (middle, shaded), the DRAM energy usage (middle,
  light), and the overhead of the Delay-on-Squash tracking structures (top,
  shaded).
  }
  \label{fig:energy}
\end{figure*}

The energy usage of the different configurations can be seen in
\autoref{fig:energy}. %
Each bar in the figure is split into four parts, representing (from bottom to
top) dynamic, static, DRAM, and overhead energy. %
The overhead contains both the dynamic and static energy usage of the squash
tracking mechanism and the Bloom filters, and is only included in the Bloom
filter configuration, as that is the only configuration where we have a
realistic hardware design. %

Overall, there are two factors that affect the energy usage in the various
configurations: Number of executed instructions and execution time. %
While we simulated all benchmarks for the same number of \emph{committed}
instructions, the number of (dynamically) executed instructions varies depending
on the amount of speculation and squashing. %
Specifically, instructions that are executed speculatively and are squashed
still require energy, but this energy is not used for useful work. %
For this reason, the dynamic energy usage of the applications is reduced in the
strict Delay-All version.

However, while restricted speculation improves the dynamic energy usage, the
static and DRAM energy usage are instead increased, because both depend on the
execution time, which is increased. %
Specifically, unless power- or clock-gated, the CPU has the same amount of
leakage regardless if instructions are executed or not. %
Then, the longer the CPU has to be active, the more the static energy
increases. %
Similarly, DRAM cells need to be refreshed at regular intervals, so the DRAM
energy increases with the execution time. %
Overall, these affects overshadow the energy gains provided by the reduced
misspeculation, so the Delay-All version has an overall energy usage that is
$23\%$ higher than the baseline. %

In contrast, Delay-on-Squash with a perfect filter, has identical
characteristics as the baseline. %
The performance difference between the two versions is very small, so there are
no significant differences in the static and DRAM energy, and speculation is not
restricted enough to lead to any energy usage reductions. %
As we are not modelling the filter mechanism, there is no overhead, but it is
important to note that storing and accessing all the squashed instructions in
the window of speculation would introduce very large overheads. %

On the other hand, the Bloom filter version has a $4\%$ energy overhead over the
baseline, out of which half (2 percentage points) is due to the increase in
execution time and half is due to the overhead of the mechanism. %
This would lead to an overall increase in the energy overhead, as the gains (due
to the reduction in the execution time) would be overshadowed by the increased
overhead due to the larger structures. %

\section{Related Work}
\label{sec:related-work}

% \cs{Add the papers mentioned in the HPCA reviews}
% \cs{InvarSpec}
% \cs{InvarSpec and Kim's work: They can help, BUT with MicroScope you might
% amplify \emph{correct} execution, so InvarSpec or Kim might say that it's safe,
% even though it is not.}
% \cs{Speculative side-channel defences explicitly want to stop INCORRECT execution, not
% CORRECT speculation, even if it leads to squashing etc. So they might not work
% with Delay on Squash.}

\fakeheader{Related Enclave Attacks and Defences}

Intel's SGX is a popular target for attacks, presumably due to the popularity of
the platform. %
Controlled-channel attacks\cite{xu_controlled-channel_2015} and Sneaky Page
Monitoring (SPM)~\cite{wang_leaky_2017} are two side-channels aimed at SGX that
take advantage of page handling, much like MicroScope does. %
However, instead of using page faults to trigger replays of the attack code,
they instead use the fact that the page management is delegated to the OS to
monitor the access patterns of the victim directly and leak information. %
As the OS has full control over the page system, these two attacks can have low
to zero noise and do not need to be repeated several times, so no
microarchitectural replay is necessary. %
CacheZoom~\cite{moghimi_cachezoom_2017} is another attack that uses fine-grained
control to perform a side-channel but only targets cache side-channels. %
As per our threat model (\autoref{sec:threat-model}), we consider such attacks
outside the scope of our work, as when such attacks are possible there is no
reason for an attacker to employ microarchitectural replay. % instead. %
We will, however, note that there exist solutions for such side-channel attacks,
such as Klotski by Zhang et al.~\cite{zhang_klotski_2020}, but they come with
steep performance and area overheads (up to $10\times$ in execution time,
depending on the configuration, for Klotski). %
There also exist speculative side-channel attacks that abuse speculative
execution and can illegally gain access to SGX data, such as
Zombieload~\cite{schwarz_zombieload_2019} or TLBleed~\cite{gras_tlbleak_2018}. %
Such attacks are affected from the same noise issues as any other side-channel
attack, and they could be used in conjunction with microarchitectural replay. %
Finally, Skarlatos et al.~\cite{skarlatos_microscope:_2019} allude, in their
MicroScope work, to possible MicroScope defences, such as fences, speculative
defences, or page fault protection defences, but they are not evaluated in
detail. %
\cs{ASPLOS abstract}

% Controlled-channel attacks are a subcategory of side-channel attacks introduced
% by Xu et al.~, where the attacker has full
% control of the side-channel. %
% Specifically, Xu et al. describe how a malicious or compromised operating system
% can take advantage of page fault handling to deduce input-dependent control flow
% and data access patterns of the application. %
% Similarly to MicroScope, the OS sets up the virtual memory of the application
% so that accesses to specific pages of interest will trigger a page fault. %
% When the application accesses the page a page fault is triggered and the OS is
% invoked. %
% However, unlike MicroScope, the OS does not use these faults to trigger
% re-execution but simply as a side-channel. %
% By monitoring the page access pattern of the application through the page fault,
% the OS can deduce the data and control patterns of the applications and in turn,
% deduce its inputs. %
% Thus, controlled-channel attacks differ from microarchitectural replay attacks
% in two major aspects: %
% First, the page faults themselves are used non-speculatively as the
% side-channel. %
% Secondly, as the attacker has full control of the side-channel, there is very
% little to no noise and thus there is no need to use MicroScope with such
% attacks. %
% Because of these two reasons, controlled-channel attacks are outside the scope
% of our work. %

% \cs{Klotski has some good related work, includes defences and why they are not
% great.}

% \cs{Table 1 from MicroScope}

\fakeheader{Side-Channel Defences}

There are numerous works on side-channel attacks and
defences~\cite{ge_survey_2018}, particularly when it comes to cache
side-channels~\cite{lyu_survey_2018}. %
There are two main research umbrellas under which such defences generally fall
under, obfuscation or isolation.

Obfuscation-based solutions try to hide the execution patterns that lead to the
side-channel. %
Examples include prefetching data to introduce noise in the cache-access
patterns~\cite{fuchs_disruptive_2015} or oblivious RAM (ORAM) to hide memory
access patterns~\cite{zhang_klotski_2020}. %
On the other hand, isolation-based solutions try to isolate the
resources used by the application from the attacker. %
These include examples such as using cache partitioning to prevent the attacker
from interfering with the victim's cache lines~\cite{wang_new_2007,
domnitser_non-monopolizable_2012, page_partitioned_2005,
kong_deconstructing_2008}. %

Each of these solutions tries to prevent or limit the creation of specific
side-channels and comes with different costs, overheads, and limitations. %
To the best of our knowledge, there exists no published work evaluating the
total cost of implementing all the different solutions that would have been
necessary to prevent all the side-channels that microarchitectural replay
attacks, such as MicroScope, can amplify. %
Also, only perfect solutions would be effective, as otherwise even the smallest
observable side-effects can be amplified by MicroScope. %

\fakeheader{Speculative Side-Channel Defences}

Since microarchitectural replay attacks abuse speculative execution, we can also
look into speculative side-channel defences that prevent the leakage of
information while executing speculatively. %
However, there is a critical difference between microarchitectural replay and
speculative side-channel attacks: %
The latter abuse speculative execution to \emph{illegally} gains access to
information while on \emph{the wrong execution path}, while, on the other hand,
microarchitectural replay attacks can amplify \emph{the correct path} and leak
information that has been accessed \emph{legally}. %
Defences and optimizations that only try to block the wrong execution
path~\cite{kim_pact20, invarspec20, speccfi20} will not work against Microscope
or other microarchitectural replay attacks. %
Furthermore, during a speculative side-channel attack, there is a \emph{data
dependence chain} between the illegal access and the information-leaking
instructions, while in a microarchitectural replay attack the handle and the
side-channel instructions have to be independent. %
Hence, speculative side-channel defences do not always work against
microarchitectural replay attacks, while also incurring higher performance
overheads. %
For example, state-of-the-art defences that block the transmission of
speculative accessed data to potential side-channel instructions, such as
NDA~\cite{weisse_nda:_2019}, STT~\cite{yu_speculative_2019, sdo20}, and
others~\cite{barber_specshield_2019, conditional19, fustos_spectreguard:_2019,
barber20}, are ineffective in this context, because, under
microarchitectural replay, the side-channel instructions may actually be in the
\emph{correct path of execution} and \emph{can also be fed with non-speculative
data}. %
At the same time, other defences that are restricted on data dependencies, such
as InvisiSpec~\cite{yan_invisispec:MICRO2018}, Delay-on-Miss~\cite{dom19,
dom20}, and many others~\cite{ghosts, safespec_2019, Saileshwar19, revice20,
muontrap20, kiriansky_dawg:MICRO2018, taram_context-sensitive_2019}, only focus
on specific side-channels, under the assumption that not all side-channels can
be exploited as easily. %
However, the effectiveness of microarchitectural replay attacks comes precisely
from the fact that they can be used to amplify and successfully mount hard to
exploit attacks, also making these speculative side-channel defences unfit for
our purposes. %

\section{Conclusion}

Microarchitectural side-channel attacks rely on observable $\mu$-state changes
caused by the victim application under attack. %
Such observations are commonly noisy, for example, due to the extremely short
lifetime of transient $\mu$-state or due to system interference that causes
persistent $\mu$-state to change. %
To successfully launch an attack, the side-channel has to be amplified and
denoised by being repeated several times. %
This is the specific purpose of MicroScope and microarchitectural replay
attacks in general, which trap the victim application in a loop, causing it to
execute the side-channel over and over again. %
MicroScope specifically, has proven to be successful in leaking information from
secure enclaves, but microarchitectural replay attacks, in general, may have
broader applicability. %
At the same time, while these attacks do abuse speculative execution, existing
speculative side-channel defences are unable to mitigate them. %

We make the observation that such replay attacks rely on repeated squashes of
replay handles and that the instructions causing the side-channel must reside in
the same ROB window as the handles. %
Based on this observation, we propose Delay-on-Squash, which is an efficient
technique for tracking squashed instructions and preventing them from being
replayed. %
We also propose an efficient implementation of Delay-on-Squash, using Bloom
filters and speculative shadow tracking~\cite{dom19}.

We evaluate several configurations with different parameters, and we show that a
fully secure system against microarchitectural replay attacks (according to the
threat model) comes at a mere $3\%$ performance degradation when compared to a
baseline, insecure out-of-order, CPU, while also keeping the energy cost low, at
$4\%$ over the baseline. %

%\section*{Acknowledgments}
% \begin{acks}

% This work was funded by Vetenskapsr\r{a}det project 2015-05159 and by MSR fund
% XXXX. %
% The computations were performed on resources provided by SNIC through UPPMAX. %

% \end{acks}

%%%%%%% -- PAPER CONTENT ENDS -- %%%%%%%%

%%%%%%%%% -- BIB STYLE AND FILE -- %%%%%%%%
% \bibliographystyle{ACM-Reference-Format}
\bibliographystyle{IEEEtranS}
% \balance
\bibliography{defs,refs}

% Generated by IEEEtranS.bst, version: 1.13 (2008/09/30)
\begin{thebibliography}{10}
\providecommand{\url}[1]{#1}
\csname url@samestyle\endcsname
\providecommand{\newblock}{\relax}
\providecommand{\bibinfo}[2]{#2}
\providecommand{\BIBentrySTDinterwordspacing}{\spaceskip=0pt\relax}
\providecommand{\BIBentryALTinterwordstretchfactor}{4}
\providecommand{\BIBentryALTinterwordspacing}{\spaceskip=\fontdimen2\font plus
\BIBentryALTinterwordstretchfactor\fontdimen3\font minus
  \fontdimen4\font\relax}
\providecommand{\BIBforeignlanguage}[2]{{%
\expandafter\ifx\csname l@#1\endcsname\relax
\typeout{** WARNING: IEEEtranS.bst: No hyphenation pattern has been}%
\typeout{** loaded for the language `#1'. Using the pattern for}%
\typeout{** the default language instead.}%
\else
\language=\csname l@#1\endcsname
\fi
#2}}
\providecommand{\BIBdecl}{\relax}
\BIBdecl

\bibitem{muontrap20}
S.~{Ainsworth} and T.~M. {Jones}, ``Muontrap: {P}reventing cross-domain
  {Spectre}-like attacks by capturing speculative state,'' in \emph{Proceedings
  of the International Symposium on Computer Architecture}, 2020, pp. 132--144.

\bibitem{aldaya_port_2019}
\BIBentryALTinterwordspacing
A.~C. Aldaya, B.~B. Brumley, S.~ul~Hassan, C.~Pereida~Garcia, and N.~Tuveri,
  ``Port {Contention} for {Fun} and {Profit},'' in \emph{Proceedings of the
  {IEEE} Symposium on Security and Privacy}.\hskip 1em plus 0.5em minus
  0.4em\relax San Francisco, CA, USA: IEEE, May 2019, pp. 870--887. [Online].
  Available: \url{https://ieeexplore.ieee.org/document/8835264/}
\BIBentrySTDinterwordspacing

\bibitem{mehdi_ooo}
\BIBentryALTinterwordspacing
M.~Alipour, T.~E. Carlson, and S.~Kaxiras, ``Exploring the performance limits
  of out-of-order commit,'' in \emph{Proceedings of the {ACM} International
  Conference on Computing Frontiers}.\hskip 1em plus 0.5em minus 0.4em\relax
  New York, NY, USA: ACM, 2017, pp. 211--220. [Online]. Available:
  \url{http://doi.acm.org/10.1145/3075564.3075581}
\BIBentrySTDinterwordspacing

\bibitem{barber20}
K.~{Barber}, A.~{Bacha}, L.~{Zhou}, Y.~{Zhang}, and R.~{Teodorescu},
  ``Isolating speculative data to prevent transient execution attacks,''
  \emph{{IEEE} Computer Architecture Letters}, vol.~18, no.~2, pp. 178--181,
  2019.

\bibitem{barber_specshield_2019}
K.~Barber, A.~Bacha, L.~Zhou, Y.~Zhang, and R.~Teodorescu, ``{SpecShield}:
  {S}hielding speculative data from microarchitectural covert channels,'' in
  \emph{Proceedings of the International Conference on Parallel Architectural
  and Compilation Techniques}, Sep. 2019, pp. 151--164.

\bibitem{bell_lipasti:ISPASS2004}
\BIBentryALTinterwordspacing
G.~B. Bell and M.~H. Lipasti, ``Deconstructing commit,'' in \emph{Proceedings
  of the International Symposium on Performance Analysis of Systems and
  Software}.\hskip 1em plus 0.5em minus 0.4em\relax Washington, DC, USA: IEEE
  Computer Society, 2004, pp. 68--77. [Online]. Available:
  \url{http://dl.acm.org/citation.cfm?id=1153925.1154589}
\BIBentrySTDinterwordspacing

\bibitem{binkert_gem5:CANEWS2011}
N.~Binkert, B.~Beckmann, G.~Black, S.~K. Reinhardt, A.~Saidi, A.~Basu,
  J.~Hestness, D.~R. Hower, T.~Krishna, S.~Sardashti, R.~Sen, K.~Sewell,
  M.~Shoaib, N.~Vaish, M.~D. Hill, and D.~A. Wood, ``The gem5 simulator,''
  \emph{{ACM} {SIGARCH} Computer Architecture News}, vol.~39, no.~2, pp. 1--7,
  Aug. 2011.

\bibitem{10.1145/362686.362692}
\BIBentryALTinterwordspacing
B.~H. Bloom, ``Space/time trade-offs in hash coding with allowable errors,''
  \emph{Communications of the {ACM}}, vol.~13, no.~7, p. 422–426, Jul. 1970.
  [Online]. Available: \url{https://doi.org/10.1145/362686.362692}
\BIBentrySTDinterwordspacing

\bibitem{cohen2003spectral}
S.~Cohen and Y.~Matias, ``Spectral bloom filters,'' in \emph{Proceedings of the
  {ACM} {SIGMOD} International Conference on Management of Data}, 2003, pp.
  241--252.

\bibitem{domnitser_non-monopolizable_2012}
\BIBentryALTinterwordspacing
L.~Domnitser, A.~Jaleel, J.~Loew, N.~Abu-Ghazaleh, and D.~Ponomarev,
  ``Non-monopolizable caches: {L}ow-complexity mitigation of cache side channel
  attacks,'' \emph{{ACM} Transactions on Architecture and Code Optimization},
  vol.~8, no.~4, pp. 35:1--35:21, Jan. 2012. [Online]. Available:
  \url{http://doi.acm.org/10.1145/2086696.2086714}
\BIBentrySTDinterwordspacing

\bibitem{fan2000summary}
L.~Fan, P.~Cao, J.~Almeida, and A.~Z. Broder, ``Summary cache: {A} scalable
  wide-area web cache sharing protocol,'' \emph{{IEEE}/{ACM} transactions on
  networking}, vol.~8, no.~3, pp. 281--293, 2000.

\bibitem{fuchs_disruptive_2015}
\BIBentryALTinterwordspacing
A.~Fuchs and R.~B. Lee, ``Disruptive prefetching: {I}mpact on side-channel
  attacks and cache designs,'' in \emph{Proceedings of the {ACM} International
  Systems and Storage Conference}.\hskip 1em plus 0.5em minus 0.4em\relax New
  York, NY, USA: ACM, 2015, pp. 14:1--14:12. [Online]. Available:
  \url{http://doi.acm.org/10.1145/2757667.2757672}
\BIBentrySTDinterwordspacing

\bibitem{fustos_spectreguard:_2019}
\BIBentryALTinterwordspacing
J.~Fustos, F.~Farshchi, and H.~Yun, ``{SpectreGuard}: {A}n efficient
  data-centric defense mechanism against {Spectre} attacks,'' in
  \emph{Proceedings of the {ACM}/{IEEE} Design Automation Conference}.\hskip
  1em plus 0.5em minus 0.4em\relax Las Vegas, NV, USA: ACM Press, 2019, pp.
  1--6. [Online]. Available:
  \url{http://dl.acm.org/citation.cfm?doid=3316781.3317914}
\BIBentrySTDinterwordspacing

\bibitem{gandolfi+2001emf}
K.~Gandolfi, C.~Mourtel, and F.~Olivier, ``Electromagnetic analysis: {C}oncrete
  results,'' in \emph{Proceedings of the International Workshop on
  Cryptographic Hardware and Embedded Systems}.\hskip 1em plus 0.5em minus
  0.4em\relax Springer, 2001, pp. 251--261.

\bibitem{ge_survey_2018}
\BIBentryALTinterwordspacing
Q.~Ge, Y.~Yarom, D.~Cock, and G.~Heiser, ``A survey of microarchitectural
  timing attacks and countermeasures on contemporary hardware,'' \emph{Journal
  of Cryptographic Engineering}, vol.~8, no.~1, pp. 1--27, Apr. 2018. [Online].
  Available: \url{http://link.springer.com/10.1007/s13389-016-0141-6}
\BIBentrySTDinterwordspacing

\bibitem{gonzalez2010processor}
A.~Gonzalez, F.~Latorre, and G.~Magklis, ``Processor microarchitecture: {A}n
  implementation perspective,'' \emph{Synthesis Lectures on Computer
  Architecture}, vol.~5, no.~1, pp. 1--116, 2010.

\bibitem{gras_tlbleak_2018}
\BIBentryALTinterwordspacing
B.~Gras, K.~Razavi, H.~Bos, and C.~Giuffrida, ``Translation leak-aside buffer:
  {D}efeating cache side-channel protections with {TLB} attacks,'' in
  \emph{{USENIX} Association}.\hskip 1em plus 0.5em minus 0.4em\relax
  Baltimore, MD: {USENIX} Association, Aug. 2018, pp. 955--972. [Online].
  Available:
  \url{https://www.usenix.org/conference/usenixsecurity18/presentation/gras}
\BIBentrySTDinterwordspacing

\bibitem{haines1994composing}
N.~Haines, D.~Kindred, J.~G. Morrisett, S.~M. Nettles, and J.~M. Wing,
  ``Composing first-class transactions,'' \emph{{ACM} Transactions on
  Programming Languages and Systems}, vol.~16, no.~6, pp. 1719--1736, 1994.

\bibitem{safespec_2019}
K.~N. Khasawneh, E.~M. Koruyeh, C.~Song, D.~Evtyushkin, D.~Ponomarev, and
  N.~Abu-Ghazaleh, ``{SafeSpec}: {B}anishing the {Spectre} of a meltdown with
  leakage-free speculation,'' in \emph{Proceedings of the {ACM}/{IEEE} Design
  Automation Conference}.\hskip 1em plus 0.5em minus 0.4em\relax Las Vegas, NV,
  USA: ACM Press, Jun. 2019, pp. 1--6.

\bibitem{revice20}
S.~{Kim}, F.~{Mahmud}, J.~{Huang}, P.~{Majumder}, N.~{Christou}, A.~{Muzahid},
  C.~C. {Tsai}, and E.~J. {Kim}, ``{ReViCe}: {R}eusing victim cache to prevent
  speculative cache leakage,'' in \emph{Proceedings of the {IEEE} Secure
  Development Conference}, 2020, pp. 96--107.

\bibitem{kim_sgx-tor_2018}
S.~Kim, J.~Han, J.~Ha, T.~Kim, and D.~Han, ``{SGX}-{Tor}: {A} secure and
  practical {Tor} anonymity network with {SGX} enclaves,'' \emph{{IEEE/ACM}
  Transactions on Networking}, vol.~26, no.~5, pp. 2174--2187, Oct. 2018.

\bibitem{kiriansky_dawg:MICRO2018}
V.~Kiriansky, I.~Lebedev, S.~Amarasinghe, S.~Devadas, and J.~Emer, ``{DAWG}:
  {A} defense against cache timing attacks in speculative execution
  processors,'' in \emph{Proceedings of the {ACM/IEEE} International Symposium
  on Microarchitecture}.\hskip 1em plus 0.5em minus 0.4em\relax Washington, DC,
  USA: IEEE Computer Society, Oct. 2018, pp. 974--987.

\bibitem{kocher+:CRYPTO1999dpa}
P.~Kocher, J.~Jaffe, and B.~Jun, ``Differential power analysis,'' in
  \emph{Proceedings of the Annual International Cryptology Conference}.\hskip
  1em plus 0.5em minus 0.4em\relax Springer, 1999, pp. 388--397.

\bibitem{kong_deconstructing_2008}
\BIBentryALTinterwordspacing
J.~Kong, O.~Aciicmez, J.-P. Seifert, and H.~Zhou, ``Deconstructing new cache
  designs for thwarting software cache-based side channel attacks,'' in
  \emph{Proceedings of the {ACM} Workshop on Computer Security
  Architectures}.\hskip 1em plus 0.5em minus 0.4em\relax New York, NY, USA:
  ACM, 2008, pp. 25--34. [Online]. Available:
  \url{http://doi.acm.org/10.1145/1456508.1456514}
\BIBentrySTDinterwordspacing

\bibitem{speccfi20}
E.~M. {Koruyeh}, S.~{Haji Amin Shirazi}, K.~N. {Khasawneh}, C.~{Song}, and
  N.~{Abu-Ghazaleh}, ``{SpecCFI}: {M}itigating {Spectre} attacks using {CFI}
  informed speculation,'' in \emph{Proceedings of the {IEEE} Symposium on
  Security and Privacy}, 2020, pp. 39--53.

\bibitem{conditional19}
P.~Li, L.~Zhao, R.~Hou, L.~Zhang, and D.~Meng, ``Conditional speculation: {A}n
  effective approach to safeguard out-of-order execution against {Spectre}
  attacks,'' in \emph{Proceedings of the International Symposium
  High-Performance Computer Architecture}, Feb 2019, pp. 264--276.

\bibitem{li_mcpat:MICRO2009}
S.~Li, J.~H. Ahn, R.~D. Strong, J.~B. Brockman, D.~M. Tullsen, and N.~P.
  Jouppi, ``{McPAT}: {A}n integrated power, area, and timing modeling framework
  for multicore and manycore architectures,'' in \emph{Proceedings of the
  {ACM/IEEE} International Symposium on Microarchitecture}.\hskip 1em plus
  0.5em minus 0.4em\relax Washington, DC, USA: IEEE Computer Society, Dec.
  2009, pp. 469--480.

\bibitem{li_CACTI:ICCAD2011}
S.~Li, K.~Chen, J.~H. Ahn, J.~B. Brockman, and N.~P. Jouppi, ``{CACTI-P}:
  {A}rchitecture-level modeling for {SRAM}-based structures with advanced
  leakage reduction techniques,'' in \emph{Proceedings of the {IEEE}/{ACM}
  International Conference on Computer-Aided Design}.\hskip 1em plus 0.5em
  minus 0.4em\relax Washington, DC, USA: IEEE Computer Society, 2011, pp.
  694--701.

\bibitem{lyu_survey_2018}
\BIBentryALTinterwordspacing
Y.~Lyu and P.~Mishra, ``A {Survey} of {Side}-{Channel} {Attacks} on {Caches}
  and {Countermeasures},'' \emph{Journal of Hardware and Systems Security},
  vol.~2, no.~1, pp. 33--50, Mar. 2018. [Online]. Available:
  \url{http://link.springer.com/10.1007/s41635-017-0025-y}
\BIBentrySTDinterwordspacing

\bibitem{matetic+:SECURITY2017rote}
S.~Matetic, M.~Ahmed, K.~Kostiainen, A.~Dhar, D.~Sommer, A.~Gervais, A.~Juels,
  and S.~Capkun, ``$\{$ROTE$\}$: {R}ollback protection for trusted execution,''
  in \emph{Proceedings of the {USENIX} Security Symposium}, 2017, pp.
  1289--1306.

\bibitem{maurice_hello_2017}
C.~Maurice, M.~Weber, M.~Schwarz, L.~Giner, D.~Gruss, C.~A. Boano, S.~Mangard,
  and K.~R\"omer, ``Hello from the other side: {SSH} over robust cache covert
  channels in the cloud,'' \emph{Proceedings of the Network and Distributed
  System Security Symposium}, 2017.

\bibitem{mcilroy_spectre_2019}
\BIBentryALTinterwordspacing
R.~Mcilroy, J.~Sevcik, T.~Tebbi, B.~L. Titzer, and T.~Verwaest, ``Spectre is
  here to stay: {A}n analysis of side-channels and speculative execution,''
  \emph{arXiv preprint arXiv:1902.05178}, Feb. 2019. [Online]. Available:
  \url{https://arxiv.org/abs/1902.05178v1}
\BIBentrySTDinterwordspacing

\bibitem{moghimi_cachezoom_2017}
A.~Moghimi, G.~Irazoqui, and T.~Eisenbarth, ``{CacheZoom}: {H}ow {SGX}
  amplifies the power of cache attacks,'' in \emph{Proceedings of the
  International Workshop on Cryptographic Hardware and Embedded Systems},
  W.~Fischer and N.~Homma, Eds.\hskip 1em plus 0.5em minus 0.4em\relax Cham:
  Springer International Publishing, 2017, pp. 69--90.

\bibitem{osvik_cache_2006}
D.~A. Osvik, A.~Shamir, and E.~Tromer, ``Cache attacks and countermeasures:
  {T}he case of {AES},'' in \emph{Proceedings of the RSA Conference}.\hskip 1em
  plus 0.5em minus 0.4em\relax Berlin, Heidelberg: Springer, 2006, pp. 1--20.

\bibitem{page_partitioned_2005}
D.~Page, ``Partitioned cache architecture as a side-channel defence
  mechanism,'' 2005, {IACR} Cryptology {ePrint} archive.

\bibitem{Saileshwar19}
\BIBentryALTinterwordspacing
G.~Saileshwar and M.~K. Qureshi, ``{CleanupSpec}: {A}n "undo" approach to safe
  speculation,'' in \emph{Proceedings of the {ACM/IEEE} International Symposium
  on Microarchitecture}, ser. MICRO '52.\hskip 1em plus 0.5em minus 0.4em\relax
  New York, NY, USA: ACM, 2019, pp. 73--86. [Online]. Available:
  \url{http://doi.acm.org/10.1145/3352460.3358314}
\BIBentrySTDinterwordspacing

\bibitem{dom20}
C.~{Sakalis}, S.~{Kaxiras}, A.~{Ros}, A.~{Jimborean}, and M.~{Själander},
  ``Understanding selective delay as a method for efficient secure speculative
  execution,'' \emph{{IEEE} Transactions on Computers}, vol.~69, no.~11, pp.
  1584--1595, 2020.

\bibitem{ghosts}
\BIBentryALTinterwordspacing
C.~Sakalis, M.~Alipour, A.~Ros, A.~Jimborean, S.~Kaxiras, and S.~Magnus,
  ``Ghost loads: {W}hat is the cost of invisible speculation?'' in
  \emph{Proceedings of the {ACM} International Conference on Computing
  Frontiers}.\hskip 1em plus 0.5em minus 0.4em\relax New York, NY, USA: ACM,
  2019, pp. 153--163. [Online]. Available:
  \url{http://doi.acm.org/10.1145/3310273.3321558}
\BIBentrySTDinterwordspacing

\bibitem{dom19}
\BIBentryALTinterwordspacing
C.~Sakalis, S.~Kaxiras, A.~Ros, A.~Jimborean, and M.~Sj\"{a}lander, ``Efficient
  invisible speculative execution through selective delay and value
  prediction,'' in \emph{Proceedings of the International Symposium on Computer
  Architecture}, ser. ISCA '19.\hskip 1em plus 0.5em minus 0.4em\relax New
  York, NY, USA: ACM, 2019, pp. 723--735. [Online]. Available:
  \url{http://doi.acm.org/10.1145/3307650.3322216}
\BIBentrySTDinterwordspacing

\bibitem{schwarz_zombieload_2019}
\BIBentryALTinterwordspacing
M.~Schwarz, M.~Lipp, D.~Moghimi, J.~Van~Bulck, J.~Stecklina, T.~Prescher, and
  D.~Gruss, ``{ZombieLoad}: {C}ross-privilege-boundary data sampling,'' in
  \emph{Proceedings of the {ACM} {SIGSAC} Conference on Computer \&
  Communications Security}, ser. Proceedings of the {ACM} {SIGSAC} Conference
  on Computer \& Communications Security.\hskip 1em plus 0.5em minus
  0.4em\relax London, United Kingdom: Association for Computing Machinery, Nov.
  2019, pp. 753--768. [Online]. Available:
  \url{https://doi.org/10.1145/3319535.3354252}
\BIBentrySTDinterwordspacing

\bibitem{skarlatos_microscope:_2019}
\BIBentryALTinterwordspacing
D.~Skarlatos, M.~Yan, B.~Gopireddy, R.~Sprabery, J.~Torrellas, and C.~W.
  Fletcher, ``{MicroScope}: {E}nabling microarchitectural replay attacks,'' in
  \emph{Proceedings of the International Symposium on Computer Architecture},
  ser. {ISCA} '19.\hskip 1em plus 0.5em minus 0.4em\relax New York, NY, USA:
  ACM, 2019, pp. 318--331, event-place: Phoenix, Arizona. [Online]. Available:
  \url{http://doi.acm.org/10.1145/3307650.3322228}
\BIBentrySTDinterwordspacing

\bibitem{spec:cpu06}
{Standard Performance Evaluation Corporation}, ``{SPEC CPU} benchmark suite,''
  \url{http://www.specbench.org/osg/cpu2006/}, 2006.

\bibitem{taram_context-sensitive_2019}
\BIBentryALTinterwordspacing
M.~Taram, A.~Venkat, and D.~Tullsen, ``Context-sensitive fencing: {S}ecuring
  speculative execution via microcode customization,'' in \emph{Proceedings of
  the Architectural Support for Programming Languages and Operating
  Systems}.\hskip 1em plus 0.5em minus 0.4em\relax Providence, RI, USA: ACM
  Press, 2019, pp. 395--410. [Online]. Available:
  \url{http://dl.acm.org/citation.cfm?doid=3297858.3304060}
\BIBentrySTDinterwordspacing

\bibitem{WWWtor}
``The {Tor} project,'' \url{https://www.torproject.org/}.

\bibitem{kim_pact20}
\BIBentryALTinterwordspacing
K.-A. Tran, C.~Sakalis, M.~Sj\"{a}lander, A.~Ros, S.~Kaxiras, and A.~Jimborean,
  ``Clearing the shadows: {R}ecovering lost performance for invisible
  speculative execution through {HW}/{SW} co-design,'' in \emph{Proceedings of
  the International Conference on Parallel Architectural and Compilation
  Techniques}, ser. Proceedings of the International Conference on Parallel
  Architectural and Compilation Techniques.\hskip 1em plus 0.5em minus
  0.4em\relax New York, NY, USA: Association for Computing Machinery, 2020, p.
  241–254. [Online]. Available: \url{https://doi.org/10.1145/3410463.3414640}
\BIBentrySTDinterwordspacing

\bibitem{wang_leaky_2017}
\BIBentryALTinterwordspacing
W.~Wang, G.~Chen, X.~Pan, Y.~Zhang, X.~Wang, V.~Bindschaedler, H.~Tang, and
  C.~A. Gunter, ``Leaky cauldron on the dark land: Understanding memory
  side-channel hazards in {SGX},'' in \emph{Proceedings of the {ACM} {SIGSAC}
  Conference on Computer \& Communications Security}, ser. Proceedings of the
  {ACM} {SIGSAC} Conference on Computer \& Communications Security.\hskip 1em
  plus 0.5em minus 0.4em\relax Dallas, Texas, USA: Association for Computing
  Machinery, Oct. 2017, pp. 2421--2434. [Online]. Available:
  \url{https://doi.org/10.1145/3133956.3134038}
\BIBentrySTDinterwordspacing

\bibitem{wang_new_2007}
\BIBentryALTinterwordspacing
Z.~Wang and R.~B. Lee, ``New cache designs for thwarting software cache-based
  side channel attacks,'' in \emph{Proceedings of the International Symposium
  on Computer Architecture}.\hskip 1em plus 0.5em minus 0.4em\relax New York,
  NY, USA: ACM, 2007, pp. 494--505. [Online]. Available:
  \url{http://doi.acm.org/10.1145/1250662.1250723}
\BIBentrySTDinterwordspacing

\bibitem{weisse_nda:_2019}
\BIBentryALTinterwordspacing
O.~Weisse, I.~Neal, K.~Loughlin, T.~F. Wenisch, and B.~Kasikci, ``{NDA}:
  {P}reventing speculative execution attacks at their source,'' in
  \emph{Proceedings of the {ACM/IEEE} International Symposium on
  Microarchitecture}, ser. {MICRO} '52.\hskip 1em plus 0.5em minus 0.4em\relax
  New York, NY, USA: ACM, 2019, pp. 572--586, event-place: Columbus, OH, USA.
  [Online]. Available: \url{http://doi.acm.org/10.1145/3352460.3358306}
\BIBentrySTDinterwordspacing

\bibitem{xu_controlled-channel_2015}
\BIBentryALTinterwordspacing
Y.~Xu, W.~Cui, and M.~Peinado, ``Controlled-channel attacks: {D}eterministic
  side channels for untrusted operating systems,'' in \emph{Proceedings of the
  {IEEE} Symposium on Security and Privacy}.\hskip 1em plus 0.5em minus
  0.4em\relax San Jose, CA: IEEE, May 2015, pp. 640--656. [Online]. Available:
  \url{https://ieeexplore.ieee.org/document/7163052/}
\BIBentrySTDinterwordspacing

\bibitem{yan_invisispec:MICRO2018}
M.~Yan, J.~Choi, D.~Skarlatos, A.~Morrison, C.~W. Fletcher, and J.~Torrellas,
  ``{InvisiSpec}: {M}aking speculative execution invisible in the cache
  hierarchy,'' in \emph{Proceedings of the {ACM/IEEE} International Symposium
  on Microarchitecture}.\hskip 1em plus 0.5em minus 0.4em\relax Washington, DC,
  USA: IEEE Computer Society, Oct. 2018, pp. 428--441.

\bibitem{yarom_flush+_2014}
\BIBentryALTinterwordspacing
Y.~Yarom and K.~Falkner, ``{FLUSH}+ {RELOAD}: {A} high resolution, low noise,
  {L3} cache side-channel attack,'' in \emph{Proceedings of the {USENIX}
  Security Symposium}.\hskip 1em plus 0.5em minus 0.4em\relax Berkeley, CA,
  USA: {USENIX} Association, 2014, pp. 719--732. [Online]. Available:
  \url{https://www.usenix.org/conference/usenixsecurity14/technical-sessions/presentation/yarom}
\BIBentrySTDinterwordspacing

\bibitem{sdo20}
J.~{Yu}, N.~{Mantri}, J.~{Torrellas}, A.~{Morrison}, and C.~W. {Fletcher},
  ``Speculative data-oblivious execution: {M}obilizing safe prediction for safe
  and efficient speculative execution,'' in \emph{Proceedings of the
  International Symposium on Computer Architecture}, 2020, pp. 707--720.

\bibitem{yu_speculative_2019}
\BIBentryALTinterwordspacing
J.~Yu, M.~Yan, A.~Khyzha, A.~Morrison, J.~Torrellas, and C.~W. Fletcher,
  ``Speculative taint tracking ({STT}): {A} comprehensive protection for
  speculatively accessed data,'' in \emph{Proceedings of the {ACM/IEEE}
  International Symposium on Microarchitecture}, ser. {MICRO} '52.\hskip 1em
  plus 0.5em minus 0.4em\relax New York, NY, USA: ACM, 2019, pp. 954--968,
  event-place: Columbus, OH, USA. [Online]. Available:
  \url{http://doi.acm.org/10.1145/3352460.3358274}
\BIBentrySTDinterwordspacing

\bibitem{zhang_klotski_2020}
\BIBentryALTinterwordspacing
P.~Zhang, C.~Song, H.~Yin, D.~Zou, E.~Shi, and H.~Jin, ``Klotski: {E}fficient
  obfuscated execution against controlled-channel attacks,'' in
  \emph{Proceedings of the Architectural Support for Programming Languages and
  Operating Systems}.\hskip 1em plus 0.5em minus 0.4em\relax Lausanne
  Switzerland: ACM, Mar. 2020, pp. 1263--1276. [Online]. Available:
  \url{https://dl.acm.org/doi/10.1145/3373376.3378487}
\BIBentrySTDinterwordspacing

\bibitem{zhao16}
C.~{Zhao}, D.~{Saifuding}, H.~{Tian}, Y.~{Zhang}, and C.~{Xing}, ``On the
  performance of {I}ntel {SGX},'' in \emph{Proceedings of the Web Information
  Systems and Applications Conference}, 2016, pp. 184--187.

\bibitem{invarspec20}
Z.~N. Zhao, H.~Ji, M.~Yan, J.~Yu, C.~W. Fletcher, A.~Morrison, D.~Marinov, and
  J.~Torrellas, ``Speculation invariance ({InvarSpec}): {F}aster safe execution
  through program analysis,'' in \emph{Proceedings of the {ACM/IEEE}
  International Symposium on Microarchitecture}, 2020.

\end{thebibliography}
%%%%%%%%%%%%%%%%%%%%%%%%%%%%%%%%%%%%

\end{document}